\UseRawInputEncoding
\documentclass[journal,compsoc]{IEEEtran}
\ifCLASSINFOpdf

\fi
\usepackage{xcolor}
\usepackage{subfigure}
\usepackage{graphicx}
\usepackage{array}                 
\usepackage{multirow}
\usepackage{longtable}
\usepackage{epstopdf}
\usepackage{algorithm}
\usepackage{algorithmic}
\usepackage{amsmath}
\usepackage{float}
\usepackage{url}

\begin{document}
%

\title{BEHAVE: Behavior-Aware, Intelligent and Fair Resource Management for Heterogeneous Edge-IoT Systems}


\author{Ismail AlQerm\authorrefmark{1}, Jianyu Wang\authorrefmark{1}, Jianli Pan\authorrefmark{1}, and Yuanni Liu\authorrefmark{2} \thanks{\authorrefmark{1} I. AlQerm, J. Wang, and J. Pan are with the Department of Computer Science, University of Missouri-St. Louis, St. Louis,
MO, USA 63121. E-mail: {alqermi, jwgxc, pan}@umsl.edu. \authorrefmark{2} Y. Liu is with the Institute of Future Network Technologies, Chong Qing University of Posts and Telecommunications, Chong Qing, China. E-mail: {liuyn}@cqupt.edu.cn.
 
\noindent * Note: This paper has been accepted by IEEE Transactions on Mobile Computing and is currently in publishing process. All rights reserved by IEEE 2021. }
 
}

\IEEEcompsoctitleabstractindextext{%

\begin{abstract}
Edge computing is an emerging solution to support the future Internet of Things (IoT) applications that are delay-sensitive, processing-intensive or that require closer intelligence. Machine intelligence and data-driven approaches are envisioned to build future Edge-IoT systems that satisfy IoT devices' demands for edge resources. However, significant challenges and technical barriers exist which complicate the resource management for such Edge-IoT systems. IoT devices running various applications can demonstrate a wide range of behaviors in the devices' resource demand that are extremely difficult to manage. In addition, the management of multidimensional resources fairly and efficiently by the edge in such a setting is a challenging task. In this paper, we develop a novel data-driven resource management framework named \textit{BEHAVE} that intelligently and fairly allocates edge resources to heterogeneous IoT devices with consideration of their behavior of resource demand (BRD). \textit{BEHAVE} aims to holistically address the management technical barriers by: 1) building an efficient scheme for modeling and assessment of the BRD of IoT devices based on their resource requests and resource usage; 2) expanding a new Rational, Fair, and Truthful Resource Allocation (RFTA) model that binds the devices' BRD and resource allocation to achieve fair allocation and encourage truthfulness in resource demand; and 3) developing an enhanced deep reinforcement learning (EDRL) scheme to achieve the RFTA goals. The evaluation results demonstrate \textit{BEHAVE}'s capability to analyze the IoT devices' BRD and adjust its resource management policy accordingly.

\end {abstract}

}
\maketitle
\IEEEdisplaynotcompsoctitleabstractindextext
\IEEEpeerreviewmaketitle
\newcommand{\rr}{\raggedright}
\section{Introduction}
Growing Internet of Things (IoT) applications raise the demand for edge computing for data processing and application intelligence. Integrating machine intelligence, IoT, and edge computing~\cite{PAN18a} benefits various applications such as virtual reality, health informatics, secure cyberspace, and urban infrastructure. It is a promising vision to build a future ``Edge-IoT'' environment converged with machine intelligence and data-driven approaches.
Edge-IoT systems can support a massive number of diverse smart devices~\cite{iot2025} with heterogeneous applications that can be delay-sensitive or require closer intelligence.
To enable such vision, there are some significant challenges to overcome. On the one hand, a massive number of IoT devices can run applications with various resource demands and different priorities. These devices may demonstrate a wide range of possible ``behavior'' in resource demand, and exhibit abnormal or abusive behavior due to potential software/hardware malfunction or IoT network attacks~\cite{DDOS162,gridhack18}. Since there is a tremendous device variety and complex behavior patterns of resource demand, more advanced methods are needed to analyze these patterns.
On the other hand, the edge servers are expected to dynamically provide multidimensional resources (CPU, memory, and bandwidth) at geospatially distributed points and different levels of network hierarchy in a fair and efficient fashion.
These challenges together severely complicate the required edge resource management and scheduling algorithms. 
Most of the current resource allocation research either focuses on a specific application, or optimizes specific operations such as mobile offloading, migration, and orchestration~\cite{CHE16,RAN15} from energy efficiency and utilization perspective. The existing edge computing schemes neither consider resource allocation with influence of behavior of resource demand (BRD) of IoT devices, nor account for fairness in resource allocation.

In this paper, we develop a novel data-driven resource allocation framework named \textit{BEHAVE} that intelligently and fairly allocates resources to heterogeneous IoT devices with consideration of their complex BRD. To holistically address the above technical challenges, and align with \textit{BEHAVE} goals, we develop the following: 
\begin {itemize}
\item A novel scheme to efficiently model and assess various BRD for heterogeneous IoT devices based on their resource requests and usage. 
The modeling approach profiles devices' BRD in two granularities: single resource request and temporal multiple requests over certain period of time. High-quality behavior features are extracted from large-dimensional behavior data using a novel deep learning technique. 
The outputs of the BRD models are leveraged to generate BRD Index for each device (BRDI) that assesses the normality of its BRD. BRDI is exploited later for resource allocation.
 
\item A new Rational, Fair, and Truthful Resource Allocation (RFTA) model to allocate the available multi-type edge resources (memory, CPU, and bandwidth) efficiently with consideration of dynamic IoT devices' BRD. 
The model aims to optimize the IoT devices' gain with consideration of their BRD and achieve fairness in the resources allocated to IoT devices that have different BRDI values; it also support the IoT devices' truthfulness while reporting their resource demands.

\item An enhanced deep reinforcement learning (EDRL) scheme for edge resource allocation to achieve the goals of the RFTA model by formulating a novel Markov decision process (MDP) and efficiently overcomes the ``dimensionality'' problem in reinforcement learning through accurate approximation of the value function.
\end{itemize}

The rest of the paper is organized as follows. The related work is presented in Section 2. Section 3 describes the motivation of \textit{BEHAVE}, its system architecture, and the design merits of its components. The scheme for assessment of IoT devices' BRD is presented in Section 4. Section 5 presents the RFTA resource allocation model. The EDRL based resource allocation scheme is illustrated in Section 6. Section 7 presents the performance evaluation and the paper concludes in Section 8. 

\vspace{-3mm}
\section{Related Work}
Resource allocation in the context of edge computing has recently been one of the emerging topics in edge computing research \cite{TRA19, ZHAN19, ZHAO19, KHAL19}. Some resource allocation proposals aim to optimize performance objectives such as power efficiency, delay, and computation rate \cite{LEG18, ZHA19, NIU19}. The work in \cite{LI18} tackled the problem of admission control and resource allocation with optimization of the devices' utility using Lyapunov dynamic stochastic optimization approach. The authors in \cite{TRA19} studied the problem of joint task offloading and resource allocation to maximize the users' tasks offloading gains. The work in \cite{ZHAO19} investigated how to allocate edge resources for average service response time minimization. 
Machine learning including reinforcement learning (RL) ~\cite{SUTT98} techniques has also been adopted for edge resource management~\cite{ NING19, YANG18}. The work in \cite{ZENG19} exploited DRL to manage the resources at the network edge. The authors in \cite{NING19} leveraged DRL to construct an intelligent offloading system for vehicular edge computing. In \cite{YANG18}, the authors designed an intelligent agent at the edge computing node to create a real-time adaptive policy for computational resource allocation of multiple users in order to improve the average end to end reliability. The deep Q-network (DQN)  approach proposed in \cite{HUA19} is a model-free approach to efficiently manage resources at
the network edge through tasks offloading and control the resource allocation among the edge servers.
The DRLRA approach proposed in \cite{KATO19} is a smart, DRL based resource allocation scheme, which allocates computing and network resources adaptively to minimize IoT devices' service time. Both approaches do not account for integrating edge resource availability from several edge servers and resource demands from heterogeneous devices with various BRD and priorities.

Given the related work for edge computing resource allocation, none of the existing work tackles the resource allocation problem with consideration of IoT devices' BRD in allocation decision-making as they only targeted optimization of certain performance metrics. 
IoT devices' behavior analysis has been widely studied in both academia and industry, which typically involves applying various machine learning techniques to analyze data traffic and achieve IoT device behavior modeling. Behavior modeling collects, processes, profiles, and models the patterns of behavior data. 
Specific behavior modeling methods were applied to various IoT management tasks, including device identification \cite{THA19} and intrusion detection \cite{JES18}, and access control \cite{SON19}.
Most of the existing work on the analysis of IoT devices' behavior are inadequate to model the complex behavior of heterogeneous IoT devices from the resource demand perspective. Moreover, the BRD of heterogeneous IoT devices needs to be profiled by considering edge computing service information such as resource usage, resource request duration, and resource request density. The modeling method should be able to measure the BRD data in various formats, length, and temporal relationship (e.g., time series and non-time series data).
\vspace{-3mm}
\section{Motivation, Architecture and Design Merits of \textit{BEHAVE} }
In this section, we present the motivation for associating the BRD of IoT devices with edge resource allocation. Moreover, we describe the architecture of \textit{BEHAVE} and the merits of the design of its components. 
\vspace{-3mm}
\subsection {Motivation of \textit{BEHAVE}}
\begin {figure}[!ht]
\vspace{-2mm}
\footnotesize
\includegraphics[width=3.4in, height=1.2in]{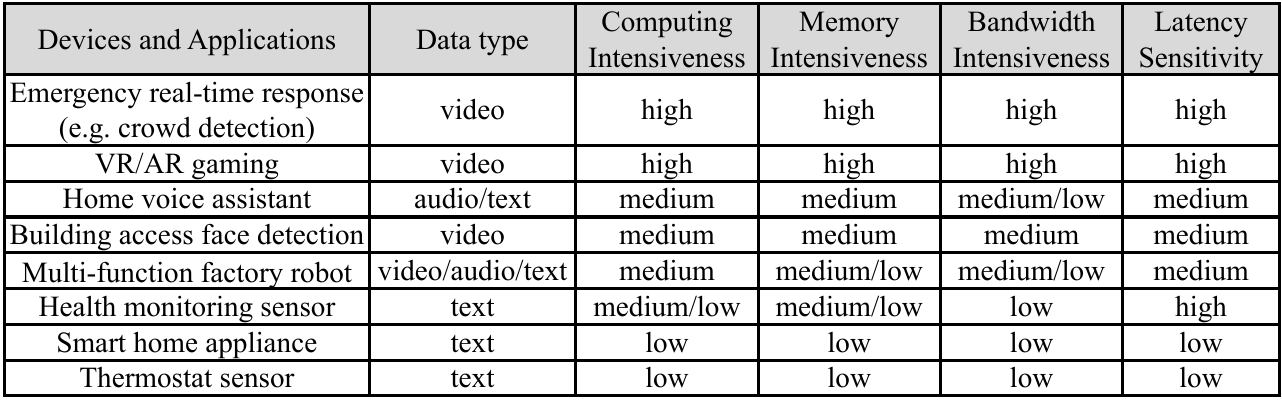}
\vspace{-3mm}
\caption{\small Edge-IoT applications and their characteristics.}
\label{fig:apps}
\vspace{-3mm}
\end{figure}

In the Edge-IoT systems, there are heterogeneous IoT devices with various demands for edge resources to 
process their tasks. Some typical examples are shown in Fig~\ref{fig:apps}. On the one hand, considering the significant impact of the irregular BRD of certain devices over the limited edge resources, it is necessary for the edge resource management framework to associate the resource allocation decisions with the devices' BRD, such that the devices with normal BRD are rewarded and the ones with irregular BRD are restricted from resource access. Without BRD monitoring, the edge computing service will be subjected to resource abuse and scheduling disorders. For example, compromised IoT devices such as botnet may launch DDoS attacks on the edge servers by flooding malicious requests to their service APIs. Malfunctioned IoT devices can also exhibit irregular BRD due to random glitches in their resource demand.
On the other hand, the edge resources are limited and are expected to be dynamically allocated at distributed points and different levels of network hierarchy. Edge resources are distributed over multiple edge servers which may have different resources utilization levels. This severely complicates the resource allocation and scheduling algorithms. Therefore, it is important to develop  resource allocation systems such that IoT devices acquire resources based on their BRD and edge resources are utilized in the most efficient manner. 
Resource allocation in such context must not only be tied to performance metrics but also support rationality through achieving certain gain for the IoT devices that exhibit normal BRD. In addition, being fair in resource allocation among devices with variable BRD and encouraging these devices to be truthful in resource demand are essential requirements such that the limited resources of the edge are not wasted.
In \textit{BEHAVE}, we associate the edge resource allocation with the BRD of IoT devices. For example, devices that request large amount of resources for applications' tasks processing without justification as their applications do not require having these resources, will be restricted. However, restriction by simply blocking these devices may not be the best practice as not all the devices exhibit the same level of irregular BRD. Some of them may have irregular BRD temporarily as they suffer an instant glitch in their operation. 
Therefore, we adopt the idea of continuous monitoring the devices BRD and control the edge resource allocation accordingly.
The Edge-IoT system considered
in this paper comprises IoT devices with heterogeneous behaviors, different
resource demands and various priorities. The rate at which IoT devices request
resources and the resource capacity at the edge change in a dynamic fashion at different time
slots. Thus, DRL ~\cite{LAVE18} is exploited in \textit{BEHAVE} to generate resource allocation actions as it is capable of modeling complex problems such as resource allocation in Edge-IoT.
Moreover, DRL does not require any prior knowledge about the Edge-IoT system model to learn certain resource allocation policy. DRL suits such resource allocation problem where the system is dynamic and long term reward is sought.

\vspace{-2.5mm}
\subsection{ Architecture of \textit{BEHAVE}}

\begin{figure}[!ht]
\footnotesize
\vspace{-2mm}
\includegraphics[width=0.49\textwidth, height=0.18 \textheight]{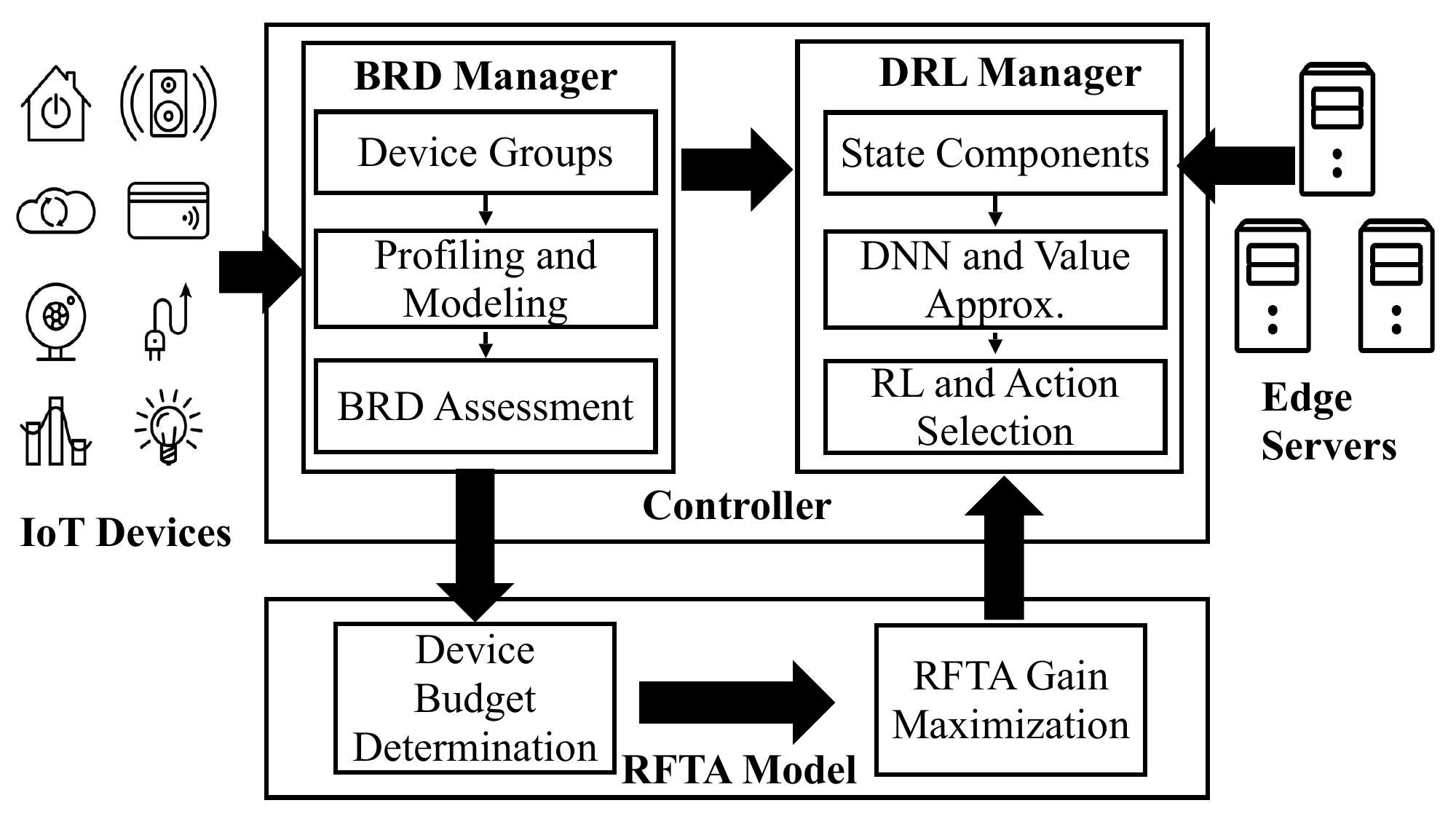}
\vspace{-6mm}
\caption{\small \textit{BEHAVE} Architecture}
\label{fig:arch}
\vspace{-6mm}
\end{figure}

\textit{BEHAVE} architecture is presented in Fig.~\ref{fig:arch}. The key components include \textit{BEHAVE} controller, the RFTA Model, the IoT devices, and the edge servers. These components coordinate with each other to
  achieve the ultimate resource allocation that maximizes the gain of IoT devices according to their exhibited BRD. The IoT devices initiate resource requests. The edge servers provide CPU, memory and bandwidth resources. Both the IoT devices and the edge servers share their state information with the managers located at the controller. For example, the IoT devices send their resource requests and the edge servers provide information about the available resources, resources usage logs and servers' capacity. The management core of \textit{BEHAVE} consists of the controller and the RFTA model which are located at the edge. The controller includes the BRD manager and the EDRL manager. The BRD manager collects devices' BRD data. Then, it creates profiles for the devices BRD, models their BRD patterns based on the built profiles, and assesses the devices' real-time BRD. We develop a novel technique that collects devices' BRD data and characterizes the interactions between the edge servers and the IoT devices. The BRD pattern of each device in the normal operation status is modeled to identify devices with irregular BRD. The identification task is formulated as an unsupervised one-class classification (OCC) problem to detect the outliers of the BRD baseline without any labeling overhead.
The EDRL manager receives the state information from the edge servers and the devices' BRD information, and formulates its MDP. Then, it runs EDRL to generate the edge resource allocation policy that aligns with the RFTA  model optimization objectives and constraints. The RFTA model aims to maximize the IoT devices' gain with efficient resource utilization under certain BRD related constraints. 
\vspace{-4mm}
\subsection {Design Merits of \textit{BEHAVE}}
In this subsection, we presents the design merits of the main components presented in \textit{BEHAVE} architecture including BRD management, RFTA model, and EDRL resource management. 
For BRD profiling, we define two granularities to profile the BRD of devices from short-term and long-term views: 1) single resource request granularity and 2) temporal requests granularity. 
These granularities align with the following operation characteristics of IoT devices: 
1) heterogeneous devices offload various IoT applications' tasks with different resource consumption and frequency density which makes their BRD connected to different time scales; 2) many IoT devices have on-demand activity patterns changing with time because of the interaction with users. For example, smart camera installed at the entrance of a building and executes a face detection application. This application has variable resource demand as the number of people entering the building is unsteady over the time.
For BRD modeling, we propose a novel Generative Adversarial Network \cite{gan}-based Encoder-Decoder (GAN-ED) deep learning framework to generate compact feature representation from large-dimension sequences. The proposed GAN-ED has the following key advantages: 1) smaller BRD dataset where the generator ability of GAN provides oversampled new samples during the training process; 2) advanced feature representation ability benefiting from the adversarial network. Thus, GAN-ED outperforms the traditional exploratory analysis measurements and statistical models \cite{Yaa10} which are not feasible to capture the data dependency in the time series \cite{Xu19} and experience difficulties to determine the model parameters.
It is also better than the widely adopted auto-encoder \cite{AE}, which is constrained by requiring a large number of data samples and lacking model ability for long-term sequence to find the time-dependent BRD pattern.
For BRD assessment, we introduce an unsupervised one-class neural network (OCNN)-based \cite{ocnn} irregular BRD assessment approach, which is concatenated with the previous deep learning structure to craft an efficient end-to-end model and reduce the requirement of costly class labeling in OCC.

 We propose RFTA model for resource allocation such that each device receives the resource allotment that is consistent with its BRD while maintaining efficient resource utilization.
 RFTA is distinguished from the existing resource allocation models because: 1) the IoT devices are assumed to have heterogeneous BRD and they have budget to obtain resources that is determined according to their exhibited BRD. 2) The resource allocation problem is formulated as a multi-resource allocation problem as the type of resources is taken into consideration and the setting of the system model incorporates multiple edge servers with multiple resource types. 3) The allocation problem is casted as behavior oriented allocation with the goal of maximizing devices' gain constrained by their budgets. 4) the proposed model also has the following unique characteristics: i) Rationality, as each IoT device achieves certain gain $\ge 0$ that is optimized to boost the IoT system performance; ii) Fairness, as the model aims to allocate resources fairly to IoT devices with distinct BRD and variable budgets; iii) Truthfulness, since the resource requests initiated by the IoT devices should be truthful in order to maximize their gain.

The proposed EDRL scheme has three unique merits. 1) It targets optimization of the IoT system performance with consideration of devices' BRD and devices' budgets. 2) It develops a novel MDP that features a composite state. This state comprises IoT device related state information such as resource demand, budget, application priority, and state of multiple edge servers including their resource capacity and utilization. Thus, our EDRL differs from the DRL used in the current other work which focuses on allocation from one centralized edge server. 
3) It incorporates a unique mapping function between the composite state elements, which enhances the accuracy of the value function approximation and consequently, the resource allocation action quality.

\vspace{-3mm}
\section{Modeling and Assessment of BRD for IoT Devices}
This section presents the scheme functionality to achieve modeling and assessment of IoT devices' BRD. \vspace{-0.3cm}

\subsection{Modeling of IoT Device BRD}
\begin{figure}[!ht]
\centering
\vspace{-2mm}
\includegraphics[width=0.45\textwidth, height=0.18\textheight]{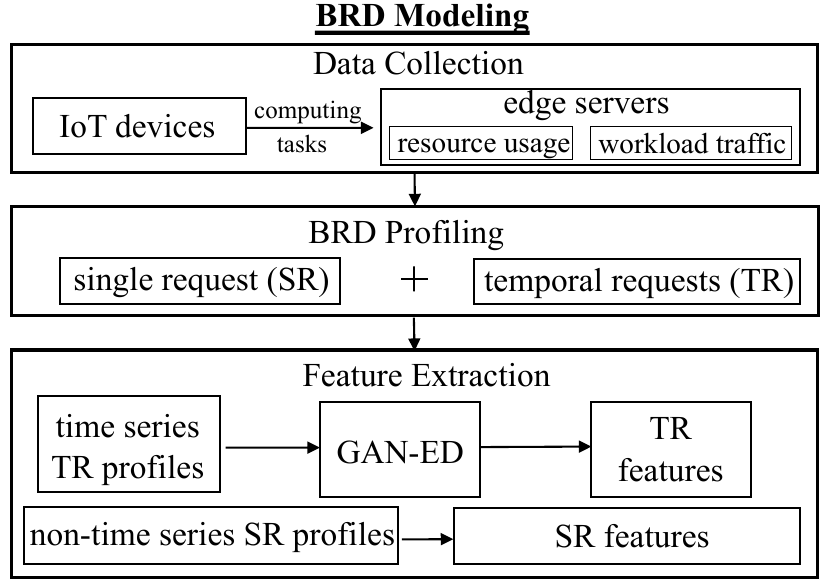}
\vspace{-2mm}
\caption{\small Modeling of devices' BRD }
\label{BRD_modeling}
\vspace{-2mm}
\end{figure}

The modeling process of each device BRD (presented in Fig. \ref{BRD_modeling}) starts with data collection in which data is collected on the edge server by logging the resource usage and the workload traffic in the normal status of the device for every resource request. Then, profiling of BRD in the two granularities engages as follows, 

\textit{1. Single Request (SR) Granularity :}
The single request granularity focuses on the device's BRD during one request received and processed by the edge servers. The BRD is profiled with multiple attributes that measure resource usage and workload traffic. The modeling in this granularity aims to identify irregular BRD in each request. The BRD profiles in this granularity include request duration, request workload size, request CPU/Mem/Disk IO usage and occupancy time, count of network connections, transport-layer protocol (e.g., TCP/UDP), and application-layer protocol (e.g., HTTPS/HTTP/MQTT/COAP).


\textit{2. Temporal Requests (TR) Granularity:}
Considering the fact that many IoT devices have on-demand activity pattern caused by the interaction between the applications and the users, the BRD of devices should also be profiled in a long-term view. Even if the current requests are identified as normal, there may exist irregular requests over a certain period of time. For example, a smart camera is not expected to send requests in a high frequency when the crowd density is low at night. 
Therefore, we propose the temporal requests granularity to profile the time-related requests. This granularity consists of three BRD profiles: the number of requests, total workload size, and resource occupancy time. The BRD data is sampled in the whole day at a fixed interval (e.g., 10 minutes). Thus, a time series is formed with a sequence of data points, where each data point is a vector of the three attributes. 

\begin{figure}[!ht]
\vspace{-2mm}
\centering
\includegraphics[width=0.45\textwidth, height=0.16\textheight]{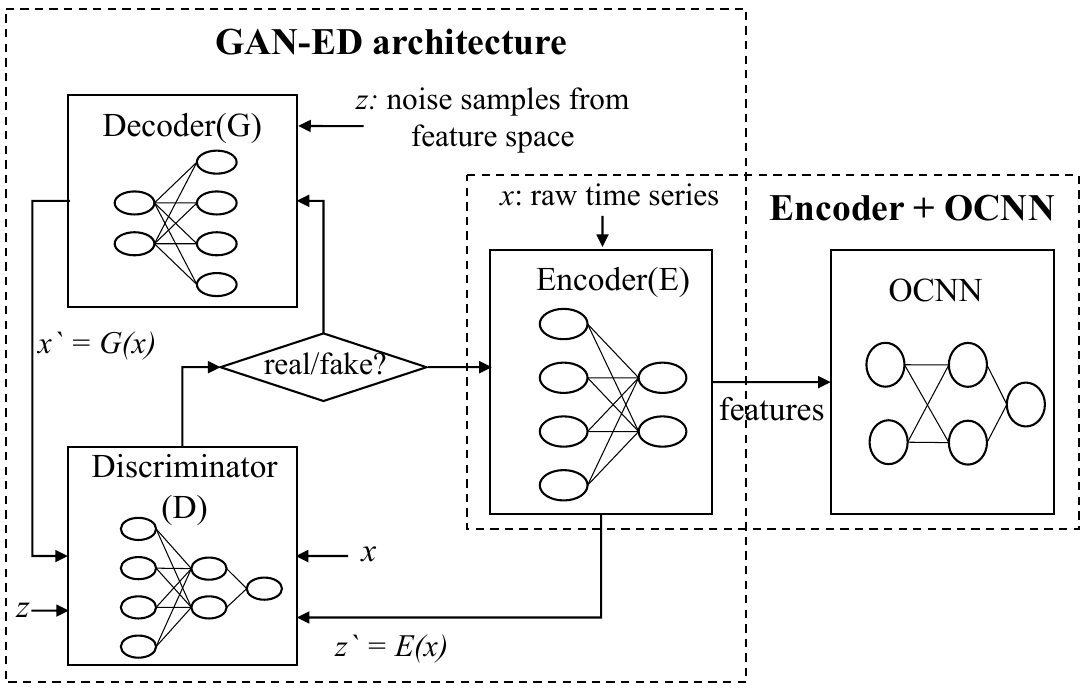}
\vspace{-3mm}
\caption{\small The architecture of GAN-ED and OCNN}
\label{GAN-ED}
\vspace{-3mm}
\end{figure}

Given the BRD profiles in TR granularity, we extract the high-quality features of BRD using GAN-ED. 
It is inspired by BiGAN \cite{bigan}, which offers the theoretical design of projecting real data into the feature space of its generative network. GAN-ED consists of an encoder network ($E$), a generative network ($G$) acting as the decoder, and a discriminator network ($D$).
In GAN-ED, we utilize an encoder ($E$) as the feature extractor that compresses a long data series into a short fixed feature vector. 
The decoder $G$ learns to map samples from an arbitrary latent distribution to the real time series distribution. The discriminator $D$ distinguishes between the real and generated time series.
In this way, $D$ advances the learning ability of both $E$ and $G$ by guiding them to learn the true data distribution from the original inputs. Moreover, we adopt LSTM \cite{lstm} neurons in the hidden layers of both $E$ and $G$ to learn the time dependencies between data points. The left part of Fig. \ref{GAN-ED} shows the model architecture and the data flow between the network components. 
Specifically, the input of $E$ is the raw time-series profiles $x$ and its output is the compressed data representation $z'=E(x)$ in the feature space. The input of $G$ comprises a sequence of noise points $z$ with the same length as the feature vector and its output is the generated time series $x'=G(z)$. The input of  $D$ consists of $x$ and the output $z'$ and $x'$ from $E$ and $G$ respectively, while the output of $D$ is the discrimination results (either valid or fake). In the model training phase, $D$ discriminates jointly original data and feature space $(x, E(x))$ versus $(G(z), z)$, where the optimization goal of $D$ is to maximize the probability of identifying $(x, E(x))$ as valid and $(G(z), z)$ as fake. On the other hand, $E$ and $G$ are jointly trained to minimize the identification probability of $(G(z), z)$. Once the training completes, the optimal $E$ and $G$ invert each other to deceive $D$. Therefore, the above training process is defined as a minimax game imposed among the three sub-networks that alternatively improve in every iteration. Let the length of input $x$ be $n$, and the similarity estimator be $Et$. The optimization function is formulated as:
\vspace{-0.2cm}
\begin{equation}
\vspace{-0.2cm}
\small
\min _{ E,G } \max _{ D }{ Et } _{ x }[log(D(x,E(x))]+{ Et }_{ z }[log(1-D(G(z),z)]
\end{equation}
Since $n$ is not fixed considering the variety of IoT devices and applications, the number of features is experimentally determined.
\vspace{-3mm}

\subsection{Assessment of BRD}
\begin{figure}[!ht]
\centering
\vspace{-2mm}
\includegraphics[width=0.45\textwidth, height=0.16\textheight]{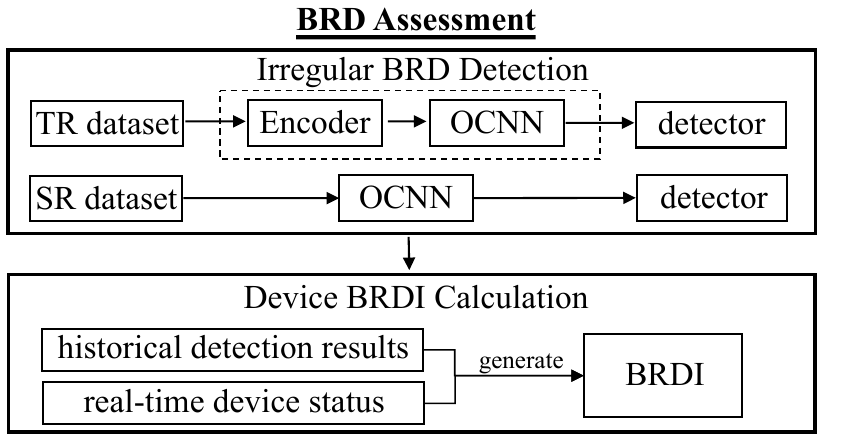}
\vspace{-3mm}
\caption{\small The scheme of devices' BRD assessment}
\label{BRD_assess}
\vspace{-3mm}
\end{figure}

In this subsection, we explain the BRD assessment process (presented in Fig.\ref{BRD_assess}) through detection of irregular BRD using the BRD datasets and the extracted features. 
\vspace{-2mm}
Furthermore, we design a longitudinal measurement for the real-time BRD of the device that uses historical detection results to generate its BRDI.
We formulate the detection task as an OCC problem, where the one class is the baseline for detection, and any outlier activity that is distinct from baseline is treated as an irregular BRD. Specifically, we construct individual detectors for the SR and TR datasets, and then measure the likelihood (in value range $[0,1]$) of a new BRD instance following the baseline. The OCNN model is adopted for detection of the irregular BRD rather than the commonly used One-class Support Vector Machine (OC-SVM) \cite{oc-svm}, because of the following unique advantages: 1) it learns the distribution of the BRD data with extra hidden neural networks where the decision boundary between normal and abnormal is highly nonlinear; 2) it models multi-dimensional features efficiently without using the computation-intensive kernel functions; and 3) it fits the previously trained Encoder network to build an end-to-end detector for the TR dataset. The OCNN model consists of one input layer, one hidden layer, and one output layer. For the SR dataset, OCNN input is the raw BRD data. For the TR dataset, however, the input layer is connected to the output layer of the encoder network, and the input is the value of the extracted features. The hidden layer is used to classify the input values according to the OCC objective. 
The output layer has sigmoid activation that calculate the probability of the BRD to be deemed as the degree of  ``normality'' in range $[0,1]$ , rather than a binary normal or abnormal classification. Let the inputs be $X$ with $N$ features, the weights of the output layer and the hidden layer be $w$ and $V$ respectively, the threshold of irregular percentage in the training dataset be $\nu$, and the decision boundary bias for normal BRD be $r$. The optimization function of the training in OCNN model is formulated as:
\vspace{-0.3cm}
\begin{equation}
\vspace{-0.1cm}
\small
 \min _{ w,V,r }{\frac { 1 }{ 2 } \left( { \left\| w \right\|  }_{ 2 }^{ 2 }+{ \left\| V \right\|  }_{ 2 }^{ 2 } \right) +\frac { 1 }{ \nu \cdot N } \sum _{ i=1 }^{ N }{ \max { \left( 0,r-\left< w,g(V{ \cdot X }_{ i } \right>  \right) -r }  } }
\end{equation}
where $\nu$ is set according to the noise percentage of the training dataset. The right part of Fig. \ref{GAN-ED} presents the integration between the encoder network in GAN-ED and the OCNN detector.

To generate the devices' BRDI, the devices are monitored and the detection of the irregular BRD is performed periodically. The BRDI is calculated jointly in the two granularities as follows,  
First, the detection results in the $SR$ granularity are recorded with time stamps. Its BRDI ($\Gamma_{SR}$)  is dynamically updated with the series of historical detection results. We assess the influence of each result according to the difference between its time and the current time. To determine the time impact, we employ Newton's Law of Cooling function~\cite{BES12}: $\Gamma_{SR}= \sum _{ t=1 }^{ T }{ { P }_{ t }\cdot{ e }^{ -\alpha t } }$, where $T$ is the number of historical records, $\left\{ { P }_{ 1 },{ P }_{ 2 },...,{ P }_{ T } \right\}$ are the corresponding BRD diagnosis records ordered from the most recent to the least recent, and $\alpha$ is the contribution parameter controlling how fast $P_{t}$ degrades.
Second, the $TR$ BRDI ($\Gamma_{TR}$) is updated based on the detection result of GAN-ED+OCNN that is performed once per day.
 Note that both $\Gamma_{SR}$ and $\Gamma_{TR}$ have the value in range $[0,1]$ indicating the BRD regularity degree.
Finally, the total $BRDI$ is jointly calculated as:
\vspace{-0.3cm}
\begin{equation}
\vspace{-0.2cm}
\Gamma = \beta_{1} \cdot \Gamma_{SR} + \beta_{2} \cdot \Gamma_{TR}
\end{equation}
where weights $\beta{1} + \beta_{2}=1$ and $\Gamma \in [0,1]$. The setting of $\beta_{1}$ and $\beta_{2}$ is determined according to the device type. If the device has no or only minor BRD change in TR granularity, $\beta_{2}$ could be close to 0. If TR BRD pattern has obvious change along the timeline, $\beta_{2}$ is given more weight. $\Gamma$ is proportional to the normality of the device BRD. \textit {BEHAVE} takes the BRDI of devices as a factor to define the devices' budgets in the subsequent resource allocation model.
\vspace{-.4cm}
\section{Rational, Fair, and Truthful Resource Allocation (RFTA) Model}
 The core of concept of RFTA is to assign different budgets for devices based on their BRD and priorities and allocate resources that match with the given budgets. For instance, the devices with low BRDI  will be given low budgets and vice versa. Each IoT device will obtain resources with certain constant costs to maximize its gain and satisfy its demand under the budget constraint. RFTA is inspired by real-world scenarios. For example, edge services can be distributed over multiple network bundles, and telecom company can grant unique budget to each bundle based on its potential revenue (subscriptions paid by the users). Similarly, IoT devices can obtain the edge resources using virtual budgets assigned according to their BRD. 
RFTA model treats IoT devices as buyers with certain budget and the edge resources are the available goods. The purchase decision is determined by the demand of the applications running on these devices and the devices' budgets which will control the range of resources they can acquire.
 \vspace{-2mm}
\subsection{System Model and Budget Determination}

In this subsection, we describe the system model and define a budget determination function for each device. The system model consists of multiple edge servers that have resources with certain capacity and different types including CPU, memory and bandwidth. In addition, the model incorporates a group of heterogeneous IoT devices that need to acquire resources to process their tasks. The communication resources are assumed to be available for the initial data transfer between the devices and the edge. As the number of devices increases, the demand for resources will increase and the complexity and overhead will escalate.
Our system aims to balance the tradeoff between performance and complexity. Thus, it assumes that the network is divided into multiple edge domains. Each edge domain consists of one controller, multiple servers and specific number of IoT devices. The number of IoT devices is matched with the network capacity. For large scale, multiple independent controllers will be needed to cover the IoT demands where each group of devices will report to one controller. Each controller is associated with multiple edge servers from the other side which ensures enough resources to serve the devices' requests. The IoT devices initially send their requests including tasks to be processed to the edge using the available network resources. The efficient allocation of network resources to connect the edge to the IoT devices is out of the scope of the paper. Once the IoT state including the resource demand is received by the controller, the controller will run EDRL scheme to find the most appropriate resource allocation action that maximize the device's gain. Thus, all the EDRL related interactions occur at the edge which is assumed to have enough  resources to handle such overhead.


The budget assigned to the IoT device is denoted as $B$. It is exploited to obtain the necessary resources and is determined according to the device's BRDI calculated in Section 4.2. 
The resource requests initiated by the IoT devices are assumed to be independent and each request aims for certain resource ratio from certain edger server. This amount of resources is called resource demand. For instance, if an IoT device requires 5 units of CPU and 3 units of memory, the resource demand will be defined as $(CPU:5, MEM:3)$. The number of successfully served requests depends on the amount of resources allocated for it. For example, if the IoT device acquires 8 units of CPU and 6 of memory, these resources will be sufficient to serve 2 requests as $\min (\frac{9}{5}, \frac{6}{3})$.
The set of edge servers, the set of IoT devices, and the set of $K$ different resource types are denoted by $\textbf{J}$, $\textbf{N}$, and $\textbf{K}$ respectively. Let $i$, $j$, and $k$ be the indexes of IoT device, edge server, and resource type. Then, the demand vector of device $i$ is defined as $ X_i = (x_{i,1}, x_{i,2}......, x_{i,K})$, where $x_{i,k}$ is the amount of resources of type $k$ requested by IoT device $i$. 
 Similarly, the amount of resources of type $k$ that is allocated to device $i$ from edge server $j$ is denoted by $y_{i,j,k}$ and the allocation vector of all resources from edge server $j$ is $  Y_{i,j} = (y_{i,j,1}, y_{i,j,2}......, y_{i,j,K}) $. The capacity of the available resources of type $k$ at the edge server $j$ is denoted by $CP_{j,k}$.
The resource allocation vector is represented by a matrix in which the row presents the resources allocated and the column is the type of resources. 
We evaluate the gain of each IoT device using RFTA Gain Function denoted by $G_{R}^i: K^{J \times K} \rightarrow K$ which maps each resource allocated to a number quantifying the device satisfaction given the resource allocated.
 It is found with respect to the number of successfully served requests $Z_i(Y_{i})$ for IoT device $i$ with $Y_i$ as the resources allocated. 
 Thus, the gain is given as,
 \vspace{-0.15cm}
\begin{equation}
\vspace{-0.2cm}
G_{R}^i (Y_i) = Z_i (Y_i) = \sum_j  Z_{i,j}(Y_{i,j})= \sum_j \min_k \frac{y_{i,j,k}}{x_{i,j,k}} \hspace{0.2cm} \forall i
\label{RFTAG}
\end{equation} 
The demand vector $X_i$ is assumed to be finite. Basically, maximizing the gain will maximize the satisfaction of IoT devices (number of successfully served requests) given the resources of different types allocated and limited by the devices' budgets.

The gain defined in (\ref{RFTAG}) is proportional to the budget $B_i$ assigned to the IoT device $i$.  
To determine the budget $B_i$, we consider three main factors: 1) the BRDI of IoT device $i$ obtained from Section 4.2; 2) the ultimate budget that should be awarded to a normal device $i$ with BRDI equals 1; and 3) the priority of the application running on the IoT device $\xi_i$. For example, the IoT applications that require real-time response are given higher priority as in Fig. \ref{fig:apps}. The ultimate budget $B_U$ is modeled as the time required by the IoT device to access certain edge resource ($T_s$). It is determined based on the amount of data $DA$ that the IoT device needs to process, the number of necessary processing units of certain resource $U_k$, and $\tau$ which is the processing factor determined according to the resource type $k$. Thus, the ultimate budget for each IoT device is given as $  B_U = \xi_i \sum_{k} T_s = \xi_i \sum_k \frac{DA}{\tau \times U_k}$.
The budget $B_i$ is given as, 
\vspace{-0.2cm}
\begin{equation}
 B_i = \Gamma_i B_U
 \vspace{-0.2cm}
 \end{equation}
where $\Gamma_i$ is the BRDI of IoT device $i$.

\vspace{-4mm}
\subsection{Problem Formulation}
In this subsection, we formulate the resource allocation problem such that IoT device $i$ uses its assigned budget $B_i$ to acquire resources from edge servers.  
The cost of resources provided from edge server $j$ is defined as $c_j = (c_{j,1}, c_{j,2},..., c_{j, K})$, each element $c_{j,k}$ represents the cost of resources of type $k$ from edge server $j$. The resource allocation for device $i$ is formulated as an optimization problem with objective of maximizing the gain $G_R^i$ as follows,
\vspace{-3mm}
\begin{equation}
\max_{Y_{i,j}} G_R^i = \sum_{j} Z_{i,j}(Y_{i,j}), \forall i \in \textbf{N} \hspace{0.2 cm}s.t.  
\vspace{-3mm}
\label {max}
\end{equation}
\begin{equation}
C.1 \hspace{0.2 cm}  \sum_{j}\sum_{k} y_{i,j,k} c_{j, k} \le B_i, \hspace{0.2 cm}
\label{cons1}
\vspace{-2mm}
\end{equation}
\begin{equation}
 C.2 \hspace{0.2 cm} \sum_{i} y_{i,j,k} \le CP_{j,k} \hspace{0.2 cm} \forall j, k, \hspace{0.2 cm} 
 \label{cons2}
 \vspace{-3mm}
 \end{equation}
 \begin{equation}
 C.3 \hspace{0.2 cm}  y_{i,j,k} > 0 \hspace{0.2 cm} \forall j, k, \hspace{0.2 cm}  C.4 \hspace{0.2 cm}  c_{j,k} \ge 0 \hspace{0.2 cm} \forall j, k
  \vspace{-2mm}
 \label{cons3}
 \end{equation}
The constraint $C.1$ in (\ref{cons1}) confirms that the amount of resources to be allocated to IoT device $i$ is controlled by its assigned budget. This 
fulfills the objective of adapting resource allocation based on the BRD of the IoT device. The constraint $C.2$ ensures that the allocated resources from certain edge server $j$ do not exceed the server $j$ capacity. In addition, the allocated resources cannot be 0 as in $C.3$. The last constraint $C.4$ indicates that the resources cannot be allocated at $c_{j,k} = 0$. This constraint also reveals that those IoT devices with BRDI $=0$ are not able to obtain any resources as they will not have budget to cover the cost. 
Our goal is to find the allocation policy that guarantees maximization of the IoT device's gain defined in (\ref{max}). 

\begin{figure}[!ht]
\includegraphics[width=0.5\textwidth, height=0.16 \textheight]{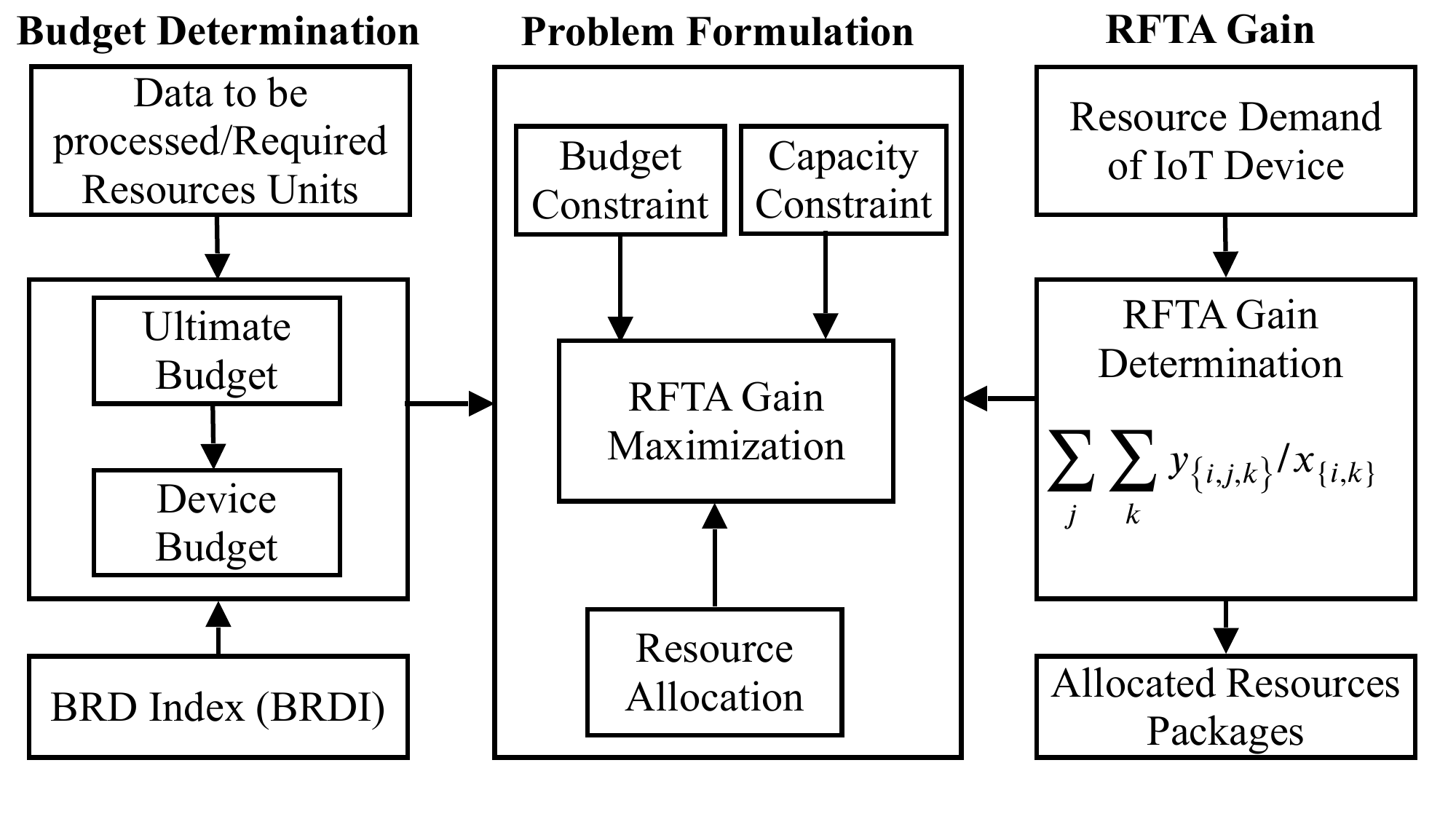}
\vspace{-9mm}
\caption{\small RFTA Model}
\label{RFTA Model}
\vspace{-3mm}
 \end{figure}

The optimization problem is formulated based on RFTA model as in Fig. \ref{RFTA Model}.
From one side, the budget of the IoT device is determined using its resource demand, the ultimate budget for such demand and the BRDI of the device. RFTA gain is exploited from the other side. The capacity and budget constraints are enforced into the optimization objective which aims to maximize the IoT devices' gain through efficient resource allocation. 

\vspace{-3mm}
\subsection{Rationality, Fairness and Truthfulness}
The formulated problem in (\ref{max}) and its associated constraints guarantee rationality, fairness, and support truthfulness as follows. 
The proposed model secures rationality for each IoT device. For example, a device $i$ with $\Gamma_i = 0$, will have budget $B_i =0$ and eventually will have $G_R^i = 0$. On the other hand, a device $l$ with $\Gamma_l > 0$, will have $B_l > 0$ and $G_R^l > 0$. This gain is valid since the cost of the allocated resources for the  device $l$ is within the confines of its budget. Hence, the proposed resource allocation model is rational as $G_R^l \ge 0$. 
 We exploit envy-freeness \cite{MOU04} feature to demonstrate that the RFTA model is fair for all IoT devices with consideration of budget variation. Envy-freeness indicates that every device is satisfied with its allocation gain and does not prefer other devices' allocation. If the budget is the same for all the devices with the same application, all the devices demonstrate envy-freeness if $Z_i(Y_i) \ge Z_i (Y_q)$, where $i \neq q$. However, since the budget in our model varies, we modified the typical definition of envy-freeness such that the allocation is envy-free if, 
\vspace{-2mm}
\begin{equation}
\vspace{-1mm}
Z_i(Y_i) \ge Z_i (\frac{B_i}{B_q}Y_q)  \hspace{0.2 cm} \forall i, q \in \textbf{N}
 \label{envy} 
\end{equation}
This is demonstrated as follows. Equation $(\ref{envy})$ can be written as, 
\vspace{-3mm}
\begin{equation}
 B_q Z_i(Y_i) \ge B_i Z_i (Y_q)  \hspace{0.2 cm} \forall i, q \in \textbf{N}
 \label{envy2} 
\end{equation}
Let $C_{i,j,k}$ be the total cost of the resources allocated with type $k$ from edge server $j$ to device $i$. According to (\ref{RFTAG}),
\vspace{-2.5mm}
\begin{equation}
B_q Z_i(Y_i) = B_q \sum_j \frac{y_{i,j,k}}{x_{i,k}} = B_q \sum_j \frac{C_{i,j,k}}{x_{i,k} c_{j,k}}, 
\vspace{-3mm}
\end{equation}
if $\psi_i \ge  \frac{1}{x_{i,k}c_{j,k}}$
\begin{equation}
B_q \sum_j \frac{C_{i,j,k}}{x_{i,k} c_{j,k}} = B_q \psi_i \sum_j {C_{i,j,k}} 
\end{equation}
\begin{equation}
B_q \psi_i \sum_j {C_{i,j,k}}  = B_q \psi_i B_i = \psi_i B_i \sum_j C_{q,j,k}
\end{equation}
\begin{equation}
 \psi_i B_i \sum_j C_{q,j,k} \ge B_i \sum_j  \frac{C_{q,j,k}}{x_{i,k}c_{j,k}} = B_i \sum_j \frac{y_{q,j,k}}{x_{i,k}} 
 \vspace{-0.1cm}
\end{equation}
\begin{equation}
B_i \sum_j \frac{y_{q,j,k}}{x_{i,k}} = B_i Z_i(Y_q)
\vspace{-3mm}
\end{equation}
Thus, the envy-freeness condition in (\ref{envy}) is satisfied.

 To justify truthfulness in RFTA model, we consider two scenarios on how IoT devices report their resource demands. In the first scenario, the IoT devices report unusual demands for resources because of their malfunction. The minimal budget assigned and consequently the limited resource allocation   will alarm the IoT users that there is a malfunction in these devices. They will only receive their fair share of resources if they truthfully report their resource demand. The second scenario is caused by the cheating IoT devices which intentionally report fraudulent resource demand. These devices will be penalized by receiving low/zero budget which makes them unable to afford the available resources. 
In both scenarios, IoT devices will have: $ \Gamma < 1$, $ B < B_U$, and $G_R \neq \max$. As a result, this guarantees that malfunctioned IoT devices will be fixed and become truthful in reporting resource demands and the abnormal devices will be encouraged to be truthful to achieve resource allocation with maximum gain.

 
\vspace{-0.3cm}
\section{Enhanced Deep Reinforcement Learning for Edge Resource Allocation}


In this section, we develop an enhanced deep reinforcement learning (EDRL) scheme for Edge-IoT resource allocation that follows MDP with RFTA gain defined in Section 5 as the optimization objective. 




\vspace{-4mm}
\subsection{MDP Formulation}
In this subsection, we formulate an MDP to maximize the RFTA gain of IoT devices defined in equation (\ref{max}). The developed MDP consists of the following components: 

\noindent \textbf{State:} 
The state is observed with each resource request arrival at epoch $t$. The state in our developed MDP is composite and includes partial state and comprehensive state elements. The partial state is defined as $ s_{j,t}= CP_j(y)$, where $ CP_j(y)$ is the available resource vector defined as $CP_j(y)= (y_{j,1}, y_{j,2}..., y_{j,K})$ at edge server $j$. Each element in this vector corresponds to the resources available of type $k$ at server $j$. The comprehensive state $s_t$ is the combination of the resource availability information for all edge servers and the device state information including budget and resource demand. It is defined as $ s_t = (CP (y), X_i, B_i)$, where $ \{CP(y)\}_{j=1}^J$ represents the aggregation of the available resource vectors $ CP_{j}(y)$. 
$X_i$ is the vector of resource demand which triggers the MDP and $B_i$ is the device's budget. 

\noindent \textbf{Action:} The action is performed when a comprehensive state transition occurs due to resource request. The action is defined as $ a_t = (Y_{i,j}= (y_{i,j,1}, y_{i,j,2}, ....., y_{i,j,k}))$ where $Y_{i,j}$ is the vector of the resource of type $k$ allocated to device $i$, and provided by edge server $j$.  The allocation action depends on not only the IoT device state, but also the edge servers state such that if $ CP_j (y)$ is saturated or the allocation cost exceeds the device's budget, allocation does not occur. Thus, the action space is defined as: 
\vspace{-0.2cm}
\begin{equation}
\vspace{-0.1cm}
A_{s_t} = 
\begin{cases}

\{0\}, \hspace{0.3 cm} Y_{i,j} > CP_j (y), $or$ 
 \hspace{0.2cm }\sum_{j} \sum_{k}y_{i,j,k} c_{j,k} > B_i \\

\{0,1\}, \hspace{0.3 cm} Y_{i,j} \le CP_j (y), $and$  
 \hspace{0.2cm } \sum_{j} \sum_{k} y_{i,j,k} c_{j,k} \le B_i \\
 
\end{cases}
\end{equation} 

\noindent \textbf{Transition Probability:}
Given the state $s_t$ and the action $a_t$, the transition to the state $s_{t+1}$ occurs as $s_t \rightarrow s_{t+1}$ in two steps. The first step determines the after-action state which is given as $ s'_{t} = f (s_t, a_t) = (CP'_j (y), X_i, B'_i)$. $ CP'_j (y)$ and $B'_i$ indicate that the capacity of edge server $j$ and the budget of IoT device $i$ respectively, are updated to reflect the resource allocation action. The second step specifies the state at $t+1$ and is defined as $ s_{t+1} = f (s'_t, X^{t+1}_{i})$. Note that the device's demand $  X^{t+1}_{i}$ is valid if there is a new resource request from device $i$ at $t+1$. The transition probability is found as $ T_R(s_{t+1}| s'_t) = {\lambda_i}/{\beta(s'_t)}$, where $\lambda_i$ is the average arrival rate of resource requests from device $i$ and $\beta(s'_t)$ is a parameter of exponential distribution to find the duration of $t$ which is the time duration of the MDP in state $s_t$ given the action $a_t$. It is evaluated according to the arrival rate of resource requests and the time given to access the allocated resources. 

\noindent \textbf{Reward Function:}
The reward function aims to fulfill the optimization objective defined in (\ref{max}) in the RFTA model. The EDRL allocation policy $\pi$ is found such that the gain of IoT devices is maximized. The policy $\pi$ is a function that determines the action $a_t = \pi(s_t)$ that the scheme selects in state $s_t$. The reward function $R (s_t, \pi(s_t))$ is derived to account for the impact of $\beta(s_t)$ according to the theory in \cite{PUTE94} as $ R(s_t, \pi(s_t)) = \rho (s_t, \pi(s_t))/ \beta (s_t)$, where $\rho (s_t, \pi(s_t))$ is found based on the number of served requests $  Z_i (Y_i)$ defined in (\ref{RFTAG}) and $R(s_t, a_t)= R'(s'_t)$ is expressed as a function of after-action state $s'_t$.
The optimal resource allocation policy of the MDP model is achieved according to after-action bellman equation as:
\begin{equation*}
V(s'_t) = \sum_{s_{t+1} \in  S} T_R (s_{t+1}| s'_t) \max_{\pi} \big( R'(s'_{t+1}) + 
\end{equation*}
\vspace{-0.2cm}
\begin{equation}
V (s'_{t+1}) - \frac{\theta}{\beta'(s'_{t+1})} \big )
\label{value}
\end{equation}
 $V (.)$ is the after-action comprehensive value function and $\theta$ is the reward rate. 
 \vspace{-3mm}
\subsection{Value Function Approximation} 

In this subsection, we illustrate the proposed novel approximation of the EDRL value function.
To formulate the value function approximation, we introduce a mapping function between the comprehensive state and the partial state after the action is selected. We denote $s'^{(m)} \in S$ as the after-action comprehensive state in its state space. $n = m(j)$ represents the index of the after-action partial state within its state space when the comprehensive state is $s'^{(m)}$. Thus, we can claim that $ 
 s'^{(m)} = \{s_j^{(m(j))}\}_{j=1}^J$.
The value function in (\ref{value}) is approximated as follows,
\vspace{-0.1cm}
\begin{equation}
\vspace{-0.2cm}
V(s'^{(m)}) = \sum_{j=1}^J \sum_{n}^{\Omega} \phi_{s'^{(n)}_j}(s'^{(m)})V_j(s'^{(n)}_j)
\label{approx}
\end{equation}
where $\Omega = |S_j|$ is the cardinality of the state space of edge server $j$ and $
V_j(s'^{(n)}_j)$ is the  value function for allocation of resources from edge server $j$ for its after-action partial state $s'^{(n)}_j$. The feature vector for the after-action comprehensive state is defined as $ \phi_{s'^{(n)}_j} (s'^{(m)}) =1$ if $
n = m(j)$ and all features of the after-action partial state $s'^{(n)}_j$ are the same for all the comprehensive states that $s'^{(n)}_j$ belongs to, that is,
\vspace{-0.1cm}
\begin{equation}
\vspace{-0.2cm}  
\{\phi_{s'^{(n)}_j}(s'^{(m)})| m \in |S|, n = m(j)\} 
\end{equation}
So far, the approximation in (\ref{approx}) is inaccurate. Hence, we use DNN to train the values $\phi_{s'^{(n)}_j}(s'^{(m)})$ and $V_j(s'^{(n)}_j)$. 
The approximation in (\ref{approx}) can be re-written as,
\begin{equation} 
 V(s'^{(m)}) = \textbf{V} \phi (s'^{(m)})
 \end{equation}
 where $\textbf{V}$ is the $ (J \times \Omega)$- dimensional value function vector with $ (j-1)\Omega +n$th element as $ V_j(s'^{(n)}_j)$. The value function in (\ref{value}) is updated with the approximation in (\ref{approx}) as follows,
 \vspace{-0.1cm}
\begin{equation*}
\vspace{-0.2cm}
\sum_{j=1}^J \phi_{s'_{j,t}} (s'_t) V_j (s'_{j,t}) = \sum_{s_{t+1} \in  S} T_R (s_{t+1}| s'_t) \max_{\pi}  \big( R'(s'_{t+1}) + 
\end{equation*}
\begin{equation}
\vspace{-0.2cm}
\sum_{j=1}^J \phi_{s'_{j,t+1}} (s'_{t+1}) V_j (s'_{j,t+1}) - \frac{\theta}{\beta'(s'_{t+1})} \big )
\label{value2}
\end{equation}
The action $a^*_t$ is selected according to the policy $ \pi^*(s_t)$ with objective of maximizing  (\ref{value2}) as follows,
\vspace{-0.2cm}
\begin{equation}
\pi^*(s_t) = \max_{\pi} \big( R'(s'_{t}) + \sum_{j=1}^J \phi_{s'_{j,t}} (s'_{t}) V_j (s'_{j,t}) - \frac{\theta}{\beta'(s'_{t})} \big )
\end{equation}
The action space $A_{s_t}$ is created and populated by the resource allocation actions $a_t = (Y_{i,j})$. 
To derive the value function vector, DNN layers' weights, and the reward rate mentioned before, we exploit stochastic gradient method under function approximation \cite{SUTT98}. Thus, the loss function is defined as,
\vspace{-0.2cm}
\begin{equation*}
\vspace{-0.2cm}
L_{t} (\textbf{V}_{t}, w^f_{t}, w_{t}) = \frac{1}{2} E \big ( \max_{\pi} \big( R'(s'_{t}) + \sum_{j=1}^J \phi_{s'_{j,t}} (s'_{t}) V_{j,t} (s'_{j,t}) - 
\end{equation*}
\begin{equation}
\phi_{s'_{j,t-1}} (s'_{t-1}) V_{j, t} (s'_{j,t-1}) - \frac{\theta_{t}}{\beta'(s'_{t})} \big ) \big)^2
\end{equation}
where $\theta_{t}$ is the reward rate up to epoch $t$. The gradient of the loss function is found using back-propagation DNN. 

\begin{figure}[!ht]
\vspace{-2mm}
\includegraphics[width=0.49\textwidth, height=0.16 \textheight]{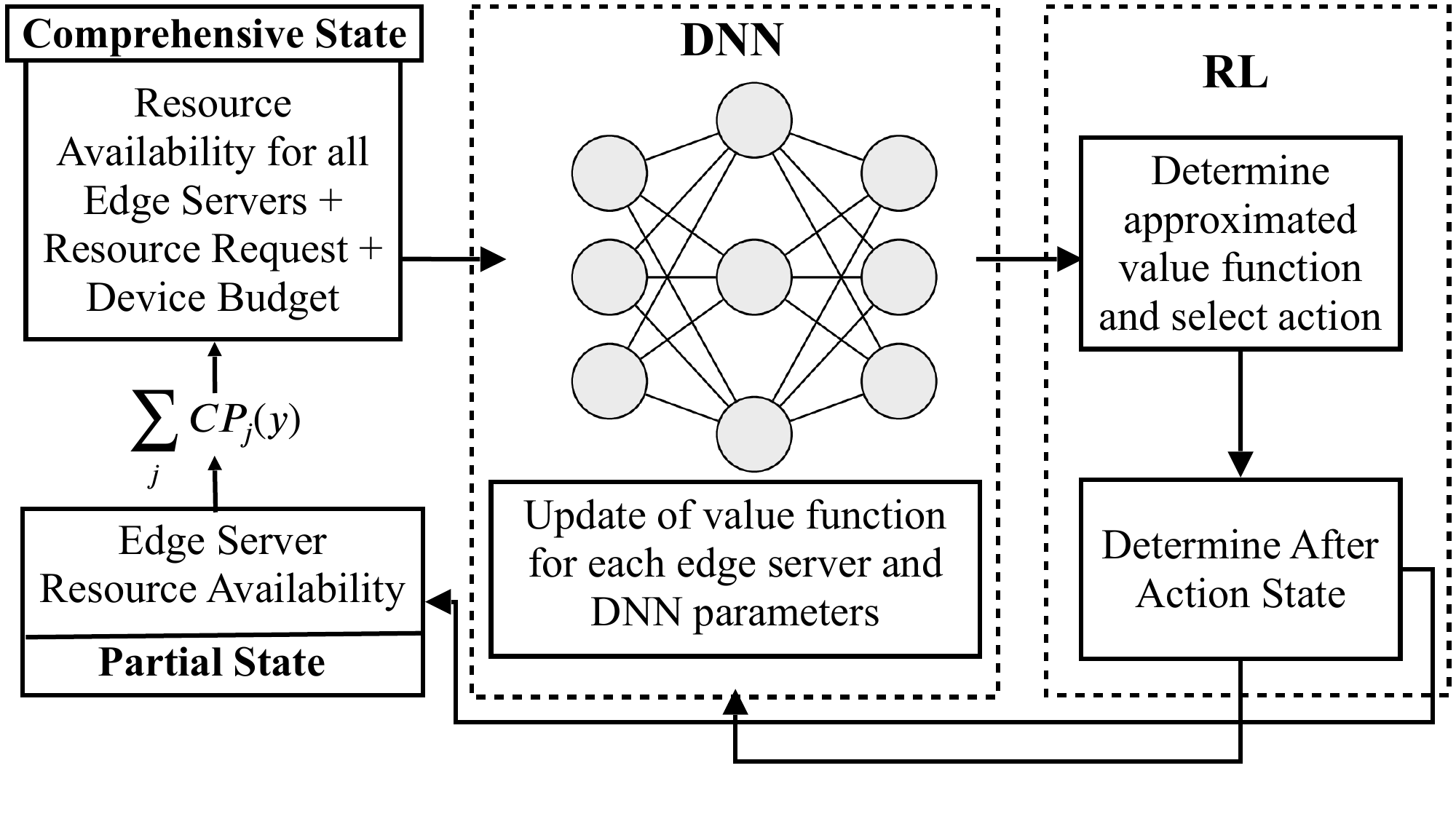}
\vspace{-10mm}
\caption{\small MDP based EDRL scheme}
\label{MDP-DRL}
\vspace{-3mm}
 \end{figure}

The MDP model integrated with EDRL is presented in Fig. \ref{MDP-DRL}. The figure shows that the state information of the IoT devices and the comprehensive state of all edge servers are exploited by the EDRL to select the most appropriate resource allocation action. DNN is used to tackle the dimensionality problem in RL through the proposed novel approximation of DRL value function. 

The inference complexity of the proposed DRL scheme executed over $T$ allocation intervals is $O (T M \log \omega + T N Y \omega)$, where $N$ and $M$ are the number of edge servers associated with the controller and the total number of IoT devices respectively. $\omega$ is the number of IoT devices requested resources at the current interval, and $Y= \sum_{j,k}y_{j,k}$ denotes the available resources at the edge servers; where $j$ and $k$ are the indexes for the edge server and resource type respectively. Note that the inference complexity per allocation interval increases linearly with the number of edge servers and IoT devices.
\section{Performance Evaluation}

We evaluate the performance of \textit{BEHAVE} in terms of the detection accuracy of the irregular BRD, detection overhead, IoT devices' gain, fairness, variance of edge server load and convergence. The evaluation studies the impact of different settings on the achieved IoT devices' gain including the number of IoT devices, the application type, the number of edge servers, and the budgets of IoT devices.
\vspace{-4mm}
\subsection{Evaluation Setup}


In the following evaluation, we simulate an Edge-IoT environment that includes 500 IoT devices and 100 edge servers with three types of resources: CPU, memory and bandwidth. These numbers are used in all the simulation unless otherwise indicated. We consider four types of IoT applications with various requirements from Fig. \ref{fig:apps}: emergency response, home voice assistant, building access face detection and health monitoring. The number of IoT devices deployed in this simulation is variable with ratio of 1/4 for each application. The resource demand (request) for each IoT device is determined according to its application. They are generated following Poisson distribution in the following ranges [0.1, 0.8] for vCPU, [0.8, 4] GB for memory and [10, 60] Mbps for bandwidth. We normalize the resource capacity of edge servers. Thus, the resource capacity of each edge server is of one unit ($CP_{j,k}= 1$  $\forall j, k$). The computing capacity of the edge servers is set between 1 GHz and 5 GHz. The average data transmission rate is distributed between 250 Mbps and 1000 Mbps. It is assumed that the budget for each IoT device is determined according to its BRDI. The budget is normalized to have value between 0 and 1. For example, the healthy IoT device will have budget equals 1 (i.e. it will have access time to edge resources that fully satisfies its demands). The resource access time at the edge servers is modeled with consideration of the following factors: edge server available  resource $ER_{j,k}$ of type $k$ such as CPU frequency, and the IoT resource demand $X_{i,k}$. The access time for resource $k$ at server $j$ is found as $T_{s} = \frac{X_{i,k}}{ER_{j,k} y_{i,j,k}} $, 
where $y_{i,j,k}$ is the amount of resource of type $k$ allocated to device $i$ from server $j$. Note that $ER_{j,k}$ varies considering different edge servers. Machine learning algorithms are executed using double Intel i7 quad core 3.4 GHz CPUs, 16 GB
Random Access Memory (RAM), and 512 GB disk. The edge servers are chosen from the set of M4 Amazon
EC2 instances \cite{amazon}. Amazon M4 instance of type M4.10xLarge includes 40 vCPU, 160 GiB of memory, and 4 GHz of bandwidth.
To evaluate the performance of \textit{BEHAVE} with respect to resource allocation, we consider the DQN scheme  proposed in \cite{HUA19}, DRLRA scheme in \cite{KATO19}, and the optimal exhaustive search for comparison.
 




\begin{figure}[!ht]
\vspace{-2mm}
\includegraphics[width=0.5\textwidth, height=0.24 \textheight]{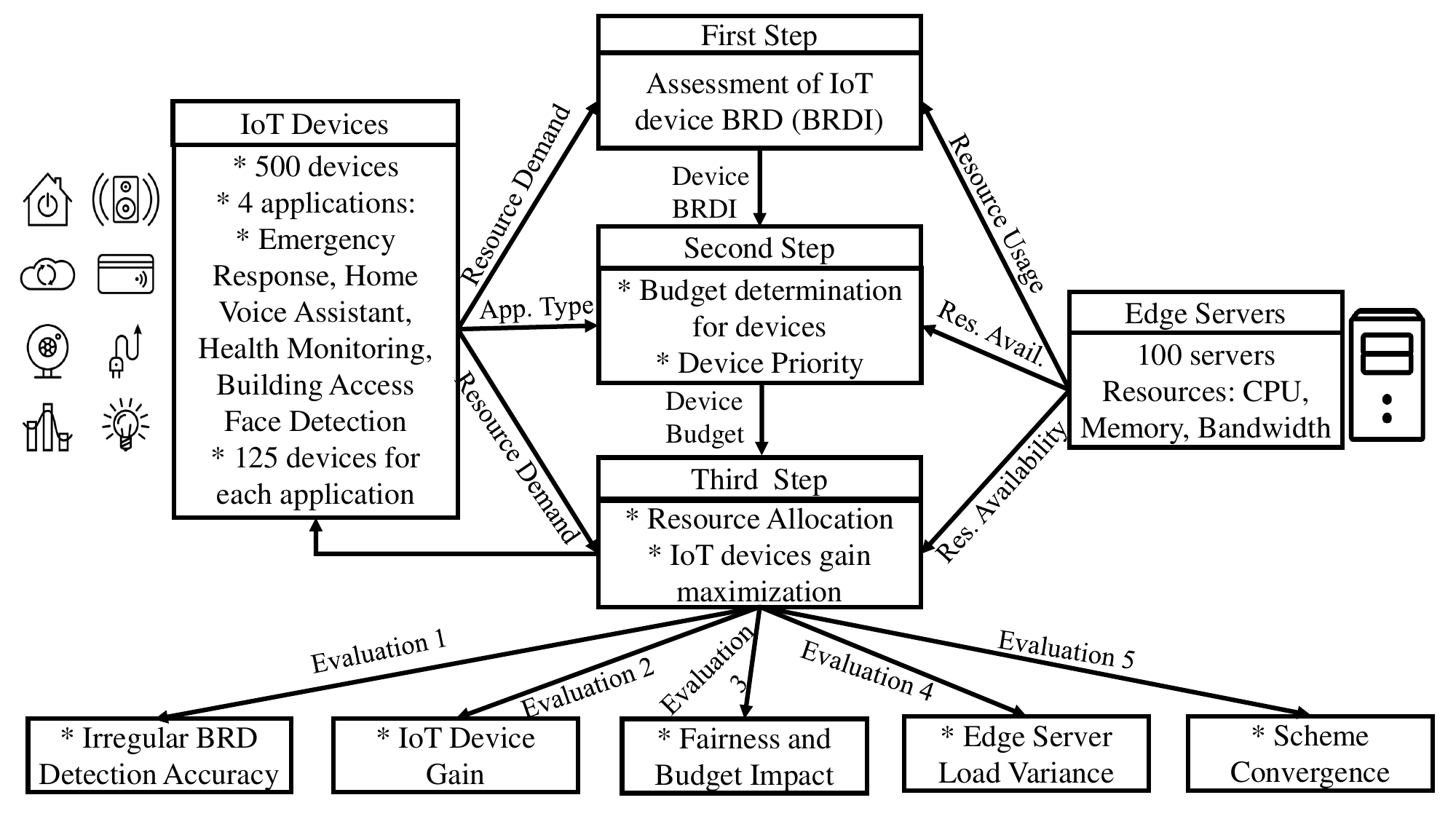}
\vspace{-5mm}
\caption{\small Evaluation setup and related steps}
\label{action}
\vspace{-4mm}
 \end{figure}

Fig. \ref{action} presents the simulation setup and the steps followed to allocate resources described as follows,

\noindent \textbf{First Step (Modeling and assessment of IoT devices' BRD):} The BRD manager in \textit{BEHAVE} learns the normal BRD patterns. When the IoT device initiates resource request to the edge, the BRD  manager dynamically synthesizes the device demands and generates its corresponding BRDI. The BRDI is updated based on current and history status of the device. 

\noindent \textbf{Second Step (Device's Budget Determination):} This step determines the budget to be assigned to each IoT device according to the following: 1) BRDI value determined at the first step; and 2) Device priority determined based on the application type. For example, a device running emergency response application will be given higher budget than a health monitoring device provided that they have the same BRDI since the emergency response application has the higher priority.

\noindent \textbf{Third Step (Resource Allocation):} This step allocates the required resources for heterogeneous devices with certain cost according to their assigned budget and the resources availability at the edge. This is achieved through the adopted MDP model and EDRL scheme with IoT devices' gain as the optimization objective.  
Various evaluations detailed in the next subsections are conducted to demonstrate \textit{BEHAVE} capabilities. 

\vspace{-0.3cm}
\subsection{Irregular BRD Detection}
\subsubsection{Detection Performance}
We implement a real Edge-IoT testbed running three typical IoT applications: building access face detection (BAFD), home voice assistant (HVA) and health monitoring (HM). These applications represent video, audio, and sensor data processing tasks on the edge nodes respectively. Thus, their behaviors are typical examples to show the modeling ability and assessment accuracy of our proposed scheme. The testbed includes edge server with 2.7 GHz CPU and 8 GB memory, and three Raspberry Pi 3 Model Bs to implement the three applications. For BAFD, the Raspberry Pi is integrated with a Pi Camera Module v2. An USB microphone is connected to the Raspberry Pi to implement HVA. We utilize a Pi sensorhat to deploy HM application. Open-source OpenCV \cite{OpenCV}, Natural Language Tool Kit \cite{language}, and Sense HAT python-based library \cite{sense} are exploited as the software platforms to process edge computing requests. Table \ref{tab:spec} presents the hardware and software platforms employed for each application.
\vspace{-0.3cm}
\begin{table}[!ht]
\footnotesize
\caption{\small Testbed Implementation Specifications}
\vspace{-3mm}
\label{tab:spec}
\centering
\begin{tabular}{|m{3.5cm}|c|p{2cm}|}
\hline
\textbf{Application} & \textbf{Hardware} & \textbf{Software} \\
\hline
Building access face detection (BAFD) &  Pi Camera  &  Open-source OpenCV\\
\hline
Home voice assistant (HVA) &  USB microphone & Natural Language Tool Kit \\
\hline
Health monitoring (HM)&   Pi sensorhat & Sense HAT\\
\hline
\end{tabular}
\vspace{-3mm}
\end{table}

\noindent To generate the irregular BRD for SR behavior granularity, we create the edge resource overuse by inserting the large sized tasks in normal traffic of a random request or occupying connection for a longer time.
For the TR granularity, we send more service requests in certain time slot. In order to evaluate the performance, F1 score \cite{F1} is adopted to compute the detection accuracy considering both the precision and recall values. 
The F1 scores from the two granularities are tested separately, in which 10 repeated experiments are made and average scores are calculated.

\begin{figure}[!ht]
\centering
\includegraphics[width=.42\textwidth, height=3.5cm]{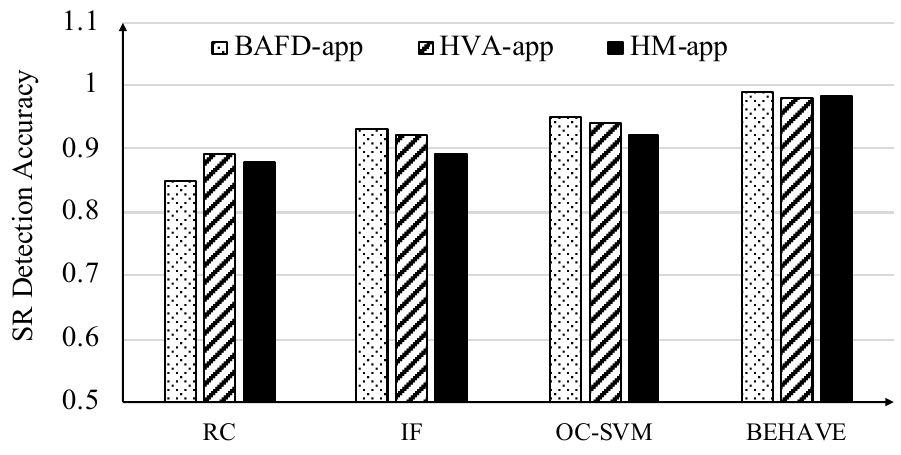}
\vspace{-3mm}
\caption{\small Irregular BRD detection accuracy comparison of models in Single-Request (SR) granularity}
\vspace{-4mm}
\label{fig:modelingPerformance_SR}
\end{figure}

\begin{figure}[!ht]
\centering
\includegraphics[width=.42\textwidth, height=3.5cm]{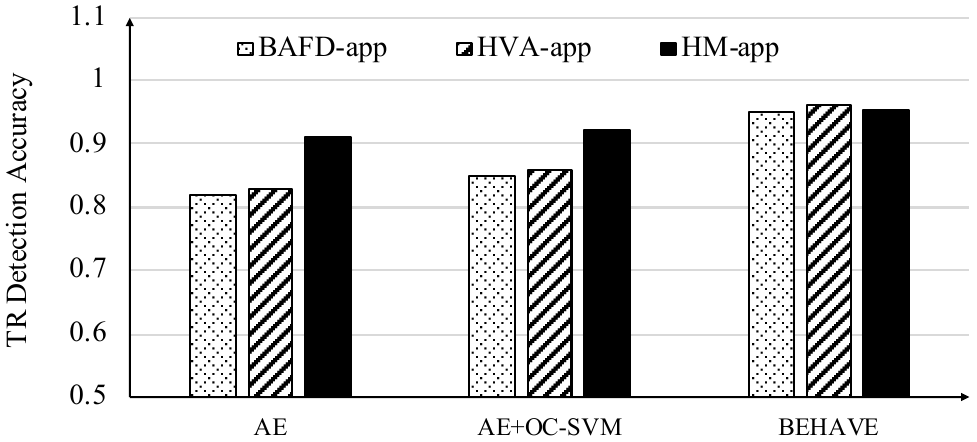}
\vspace{-3mm}
\caption{\small Irregular BRD detection accuracy comparison of models in Temporal-Requests (TR) granularity}
\vspace{-6mm}
\label{fig:modelingPerformance_TR}
\end{figure}

For the SR, we compare OCNN model in \textit{BEHAVE} to other three state-of-the-art OCC detection methods, including Robust Co-variance (RC) \cite{RC}, Isolation Forest (IF) \cite{IF}, and OC-SVM \cite{oc-svm}. 
Fig. \ref{fig:modelingPerformance_SR} shows that \textit{BEHAVE} outperforms the other models for all the three applications. For the TR, we compare GAN-ED+OCNN model in \textit{BEHAVE} to other two popular detection methods, Autoencoder (AE) \cite{AE} and integration of Autoencoder and OC-SVM (AE+OC-SVM) \cite{Sar16}. Fig. \ref{fig:modelingPerformance_TR} shows that the proposed GAN-ED+OCNN achieves better detection accuracy than the other models. We also notice that HM has significant accuracy variation using AE and AE+OC+SVM models. The reason is that HM has relatively more static behavior than BAFD and HVA.
The feature learning ability of GAN-ED for time-series BRD is a key contributor to the performance.

\vspace{-3mm}

\subsubsection{Detection Overhead}
\begin{figure*}[t]
\centering
\subfigure[\footnotesize The gain of all devices vs the number of edge servers]{\includegraphics[width=5.9cm,height=3.5cm] {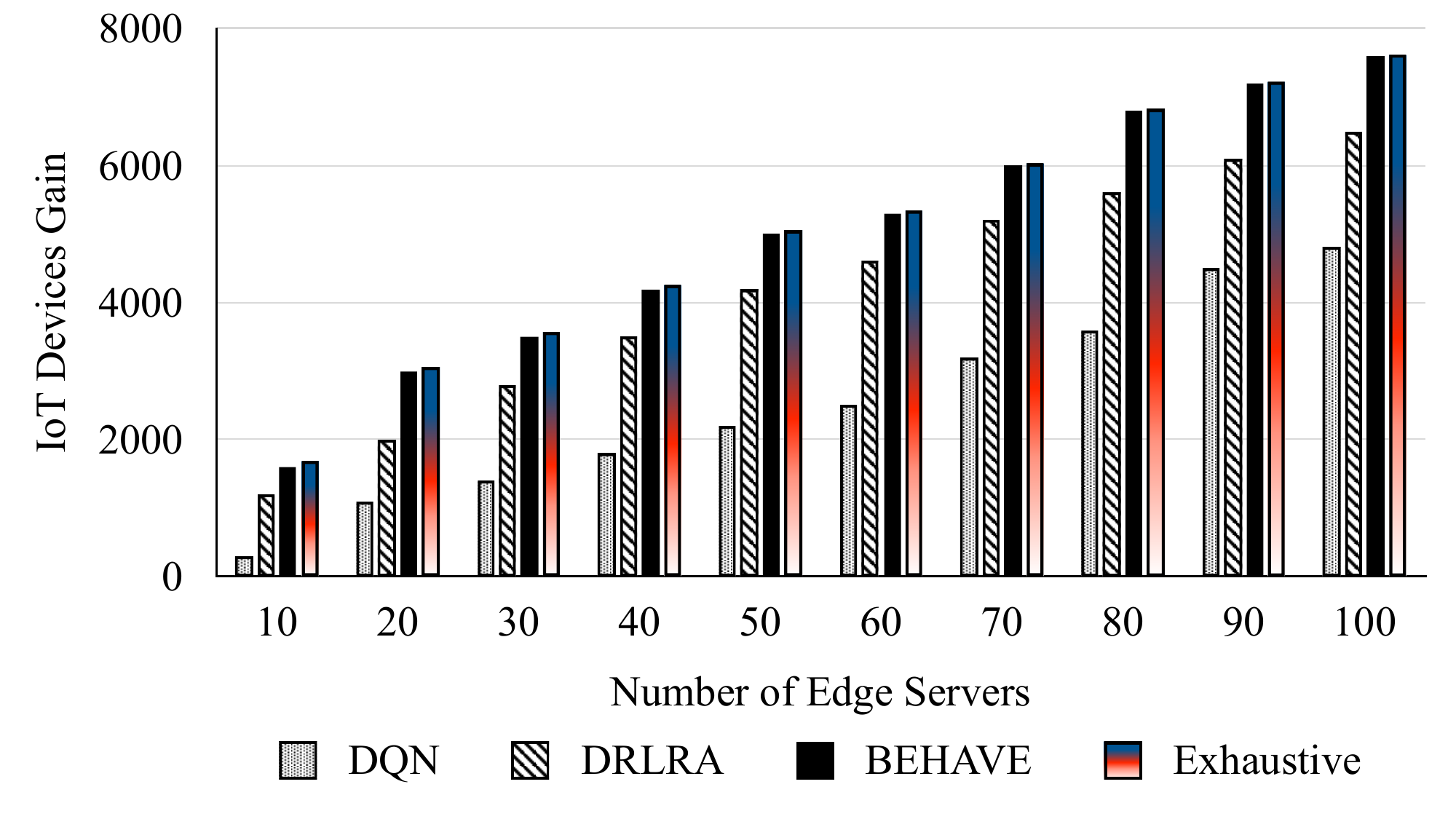} \label{gain1}\hspace{-0.1cm}}
\subfigure[\footnotesize The gain of one device vs the number of IoT devices]{\includegraphics[width=5.9cm,height=3.5cm]
{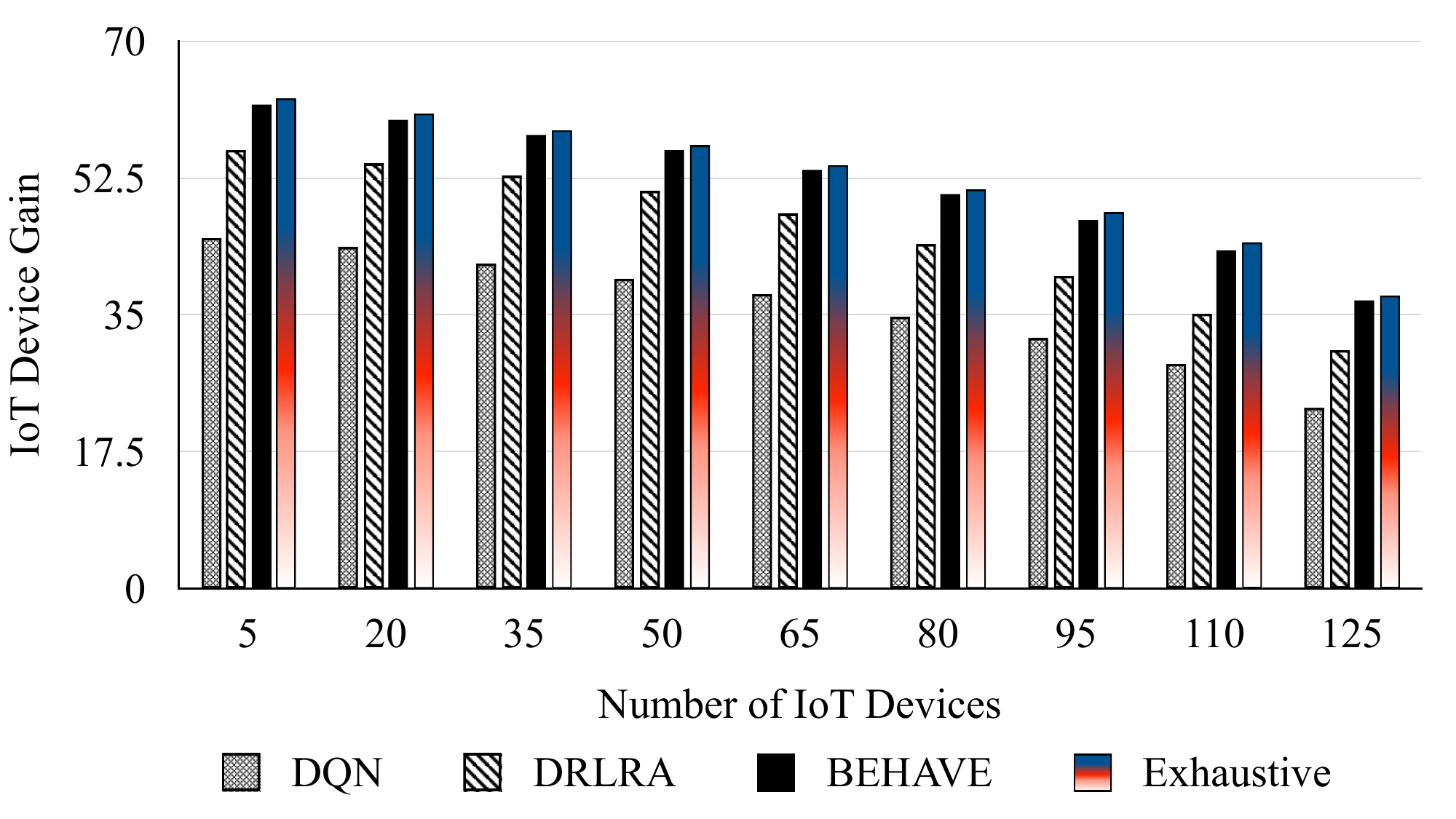} \label{gain2}\hspace{-0.1cm}}
\subfigure[\footnotesize The gain of all IoT devices vs the edge processing capacity]{\includegraphics[width=5.9cm,height=3.5cm]
{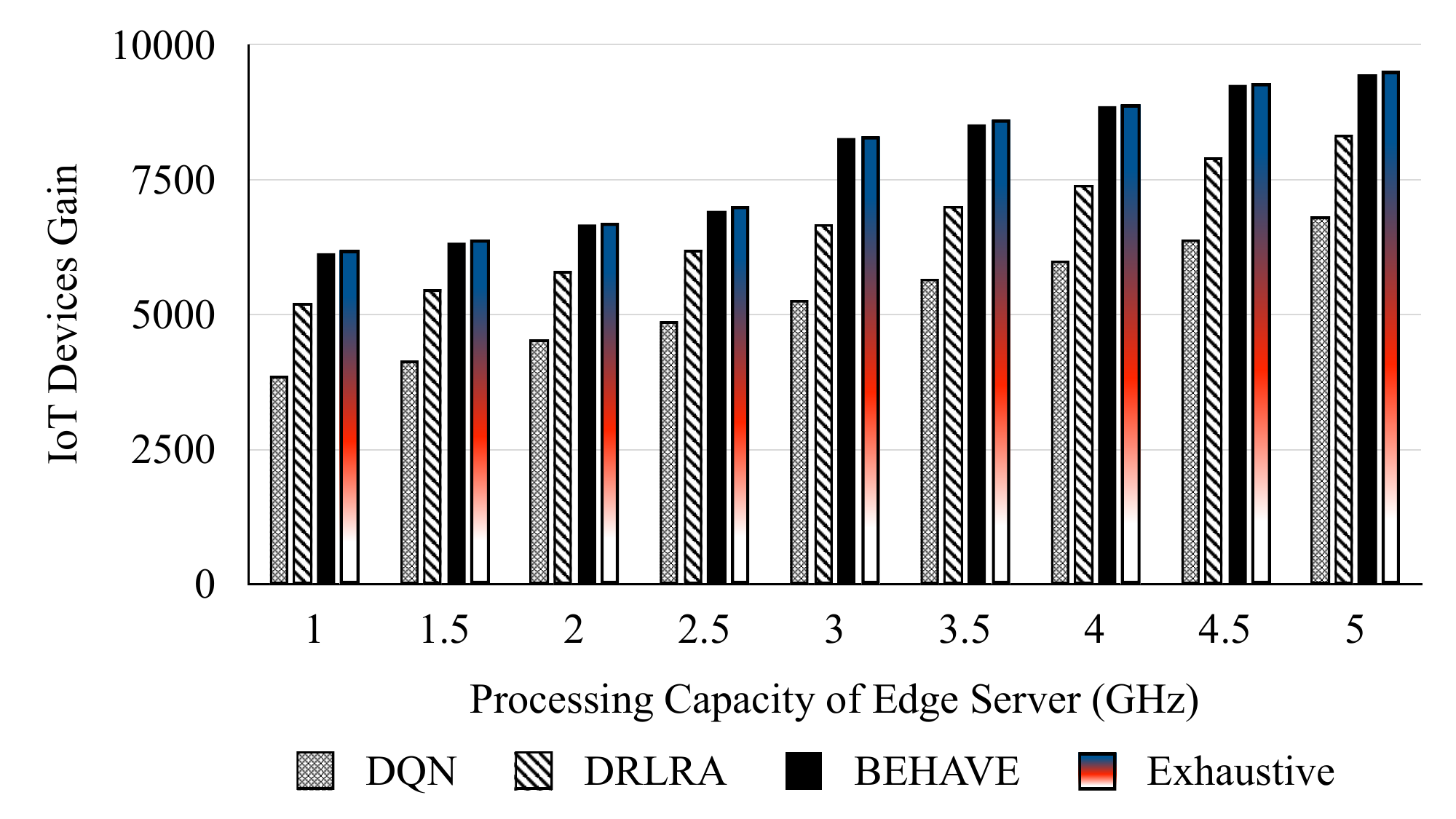} \label{gain3}\hspace{-0.1cm}}
\label{fig_WatchMap}
\vspace{- 0.1 cm}
\caption{\small IoT gain Evaluation}
\label{fig_WatchMap}
\vspace{-0.4cm}
\end{figure*}
We evaluate the overhead of BRD detection using theoretical analysis and experimental results.
For the theoretical analysis, we analyze the time complexity for SR and TR BRD assessment. 
For SR modeling, OCNN consists of three layers: input, hidden, and output layer. Let the number of features of the input layer be $N$, the weight of the hidden layer be $V$, the weight of the output layer be $W$. The time complexity for evaluating the SR BRD sample is $O(N*V*W)$.
For TR modeling, time complexity is mainly introduced by the encoder of GAN-ED. It adopts one LSTM layer to learn time dependency among data points. Let the length of input sequence be $n_{i}$, the number of memory units (each unit includes one cell, one input gate, one output gate, and one forget gate) 
be $n_{c}$, and the number of output features be $n_{o}$. The time complexity is calculated as $O(n_{i} * n_{c} + n_{c} * n_{c} + n_{c} * n_{o})$. The computation time of the encoder with a small number of inputs $n_{i}$ and $n_{o}$ is dominated by the $n_{c} *n_{c}$ factor. 
Therefore, the time complexity of the whole GAN-ED + OCNN model is $O(N*V*W + n_{c}*n_{c})$. Considering that $N>V>W$, $n_{c} <= 144$, and features number $N$ is limited in our case, the behavior assessment can be completed in polynomial time without introducing much overhead to the edge server.
For the experimental results, we implement an evaluation test to obtain the cost of BRD modeling and assessment. The detailed setting includes: 1) the training dataset of 2000 samples and each samples is a sequence with 144 data points; 
2) the training iteration is 20 epochs. We perform ten tests to get model training time (MTT), behavior detection time (BDT) using the model, and model running memory (MRM). Table~\ref{tab:detection_overhead} shows the results, which indicate the proposed scheme is 
deployable on the typical edge server at acceptable cost.
\vspace{-3mm}
\begin{table}[h]
\footnotesize
\caption{\small Detection Overhead Evaluation}
\vspace{-3mm}
\label{tab:detection_overhead}
\centering
\begin{tabular}{|l|c|c|c|}
\hline
\textbf{Model} & \textbf{MTT (s)} & \textbf{BDT (ms)} & \textbf{MRM (MB)} \\
\hline
GAN-ED &  18 - 20 s & 190 - 230 ms &  163 - 184 MB\\
\hline
OCNN &  9 - 11 s & 65 - 95 ms & 63 - 89 MB \\
\hline
\end{tabular}
\vspace{-3mm}
\end{table}
\vspace{-4mm}

\subsection{IoT Devices Gain}
The achieved gain value maps the resources allocated to a number that quantifies the IoT satisfaction given the resources allocated $G_R^i (Y_i): K^{M \times K} \rightarrow K$, where $K$ is the resource index defined in the system model.
In the following, we study the impact of the number of IoT devices, the number of edge servers, and the edge servers processing capacities on the achieved gain of the IoT devices.
Fig. \ref{gain1} shows the gain of all IoT devices vs. the number of edge servers. The figure indicates that the gain increases as there are more edge servers available to provide more resources that satisfy the demand of IoT devices.
To study the impact of the number of IoT devices on the achieved gain which mainly impacts the demand for resources from the edge, we evaluate the performance for 125 devices that run the emergency response application.
Fig. \ref{gain2} presents the achieved gain for each IoT device vs the number of IoT devices. 
It is definite that the increase in the number of IoT devices leads to a decline in the achieved gain as the demand increases and competition for resources between IoT devices escalates. \textit{BEHAVE} maintains the gain at reasonable level when the number of IoT devices is large. 
Fig. \ref{gain3} presents the IoT devices' gain as a function of the processing capacity of edge servers.
It is observable from Fig. \ref{gain3} that as the computing power of edge servers increases, the devices' gain gradually improves. Thus, the number of satisfied resources requests increases. It is also obvious that when the computing capability of the edge server is moderate, a performance gap exists between \textit{BEHAVE} and other schemes. 
We clearly notice that \textit{BEHAVE} outperforms other DRL based resource allocation schemes in the IoT devices' gain evaluation specifically at critical system settings such as small number of edge servers and large number of IoT devices. 
 In addition,  
we notice that \textit{BEHAVE} achieves IoT gain close to the optimal exhaustive search based scheme that is within the range of 1\%. Exhaustive search requires searching through all the possible resource allocation possibilities. It is impractical in the considered edge-IoT system given the fact that the search becomes complicated and consumes significant time as the network scale grows in terms of the numbers of IoT devices, edge server and edge resources.
%
\vspace{-4mm}
\subsection{Fairness and Budget Impact}
\begin{figure}[!ht]
\vspace{-2mm}
\includegraphics[width=.42\textwidth, height=3.5cm]{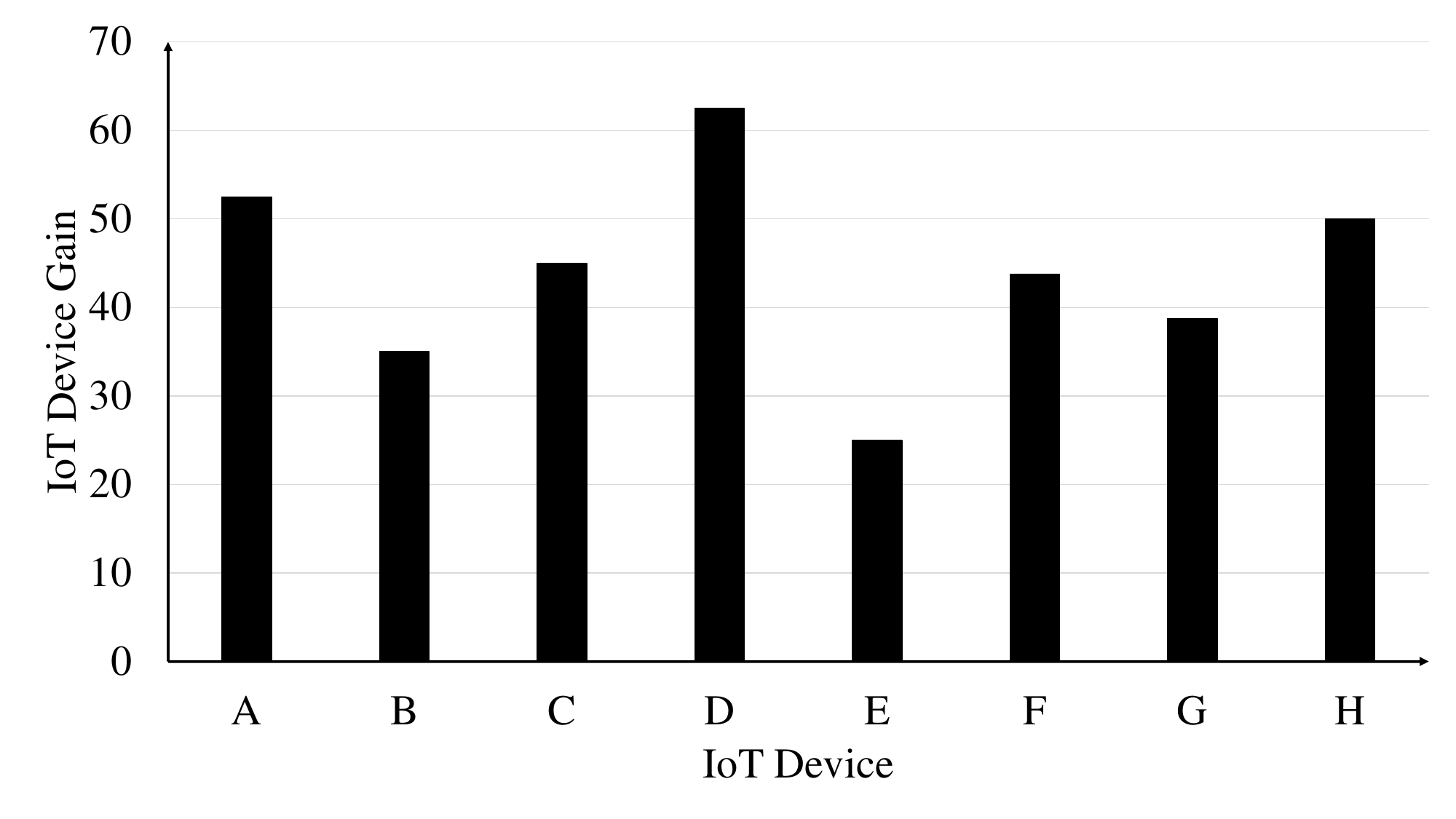}
\vspace{-4mm}
\caption {\small IoT gain with variable devices budgets} 
\vspace{-4mm}
\label{gain4}
\end{figure}

In this evaluation, we demonstrate \textit{BEHAVE} capability to maintain fairness among different IoT devices in resource allocation. In addition, we explore the impact of the variation of devices' budgets on the achieved gain. 
\begin{figure}[!ht]
\vspace{-2mm}
\includegraphics[width=.42\textwidth, height=3.5cm]{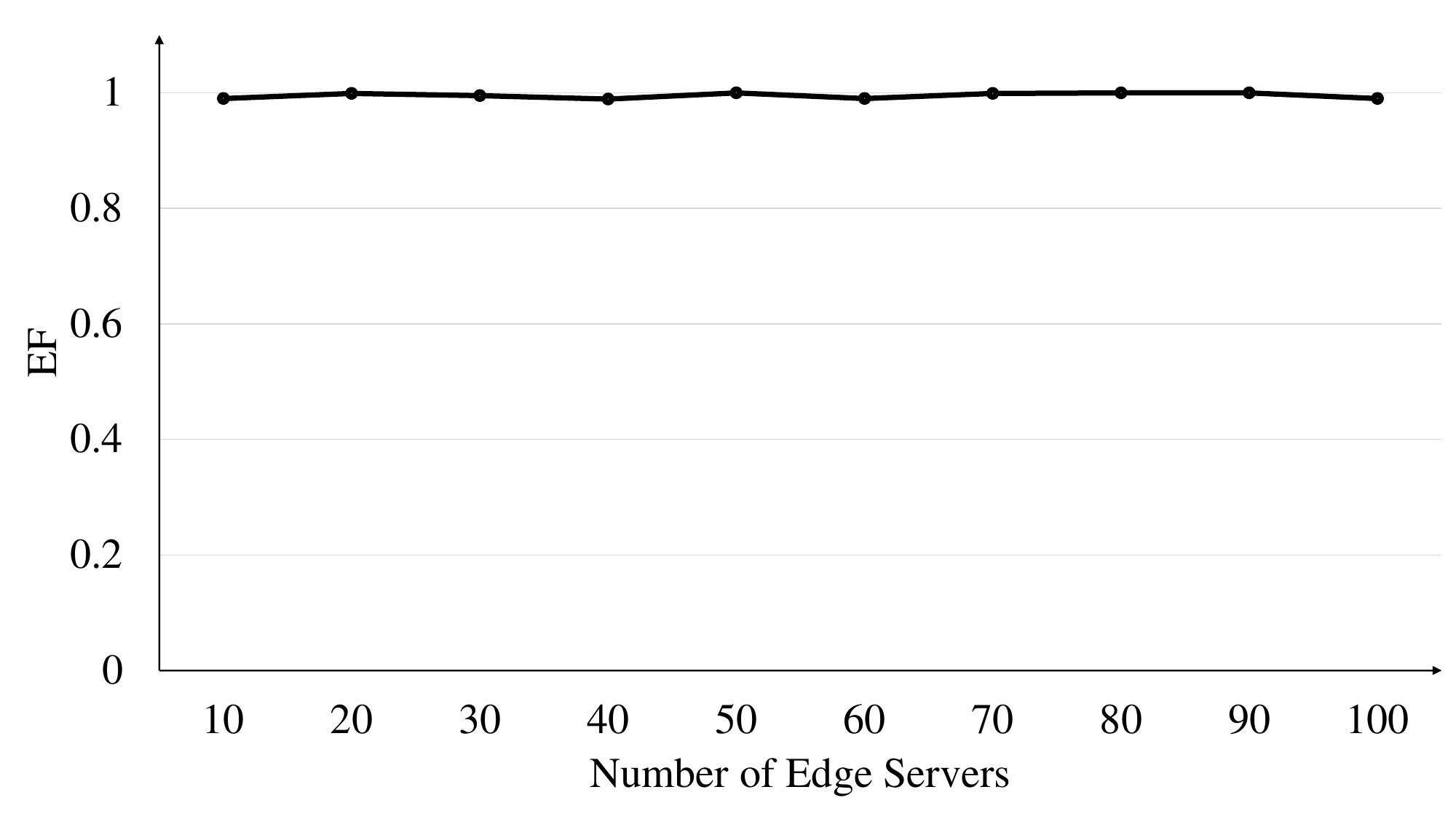}
\vspace{-4mm}
\caption {\small EF index with variable number of edge servers} 
\vspace{-2mm}
\label{EF}
\end{figure}
The fairness is evaluated using envy freeness index (EF) which is found as $EF(y) = \min_{i,q} Z_i(y_i) / Z_i (\frac{B_i}{B_q}y_q)$.
Allocation is envy-free if the EF index equals one. Fig. \ref{EF} shows that \textit{BEHAVE} records an EF equals approximately one which confirms the proved envy free property in Section 5.3. 
To study the impact of budget on the gain of the IoT devices, we selected a group of 8 IoT devices where each 2 devices run one of the applications stated in the evaluation setup. IoT devices, their applications and the BRD indexes are given in Table 3. We notice that device D which runs emergency response  application and records the highest BRDI of 0.98, achieved the highest gain as it has the highest budget. The gain for each IoT device is presented in Fig. \ref{gain4}.
\begin{table} [!ht]{
\caption{\small IoT devices with various applications and BRDIs}
\vspace{-0.3cm}
\centering
\scalebox {0.85}{
\begin{tabular}{|m{2.5 cm}|p{4cm}|p{2cm }|c}
\hline
\textbf {Device} & \textbf {Application}& \textbf {BRDI}\\
\hline
Device A & Health Monitoring & 0.98 \\
Device B  &Health Monitoring & 0.75  \\
Device C & Emergency Response & 0.85  \\
Device D & Emergency Response & 0.98 \\
Device E & Home Voice Assistance  & 0.65 \\
Device F & Home Voice Assistance  & 0.89 \\
Device G & Building Access Face Detection & 0.78 \\
Device H & Building Access Face Detection & 0.95 \\
\hline
\end{tabular}}}
\end{table}
This evaluation clearly demonstrates the budget impact on the gain achieved. In addition, it shows that \textit{BEHAVE} is effective in capturing the application priority in resource allocation as even if two devices have the same BRDI but with different applications, they achieve different gain.


\subsection{Resource Allocation Balancing}
\begin{figure}[!ht]
\includegraphics[width=.47\textwidth, height=3.5cm]{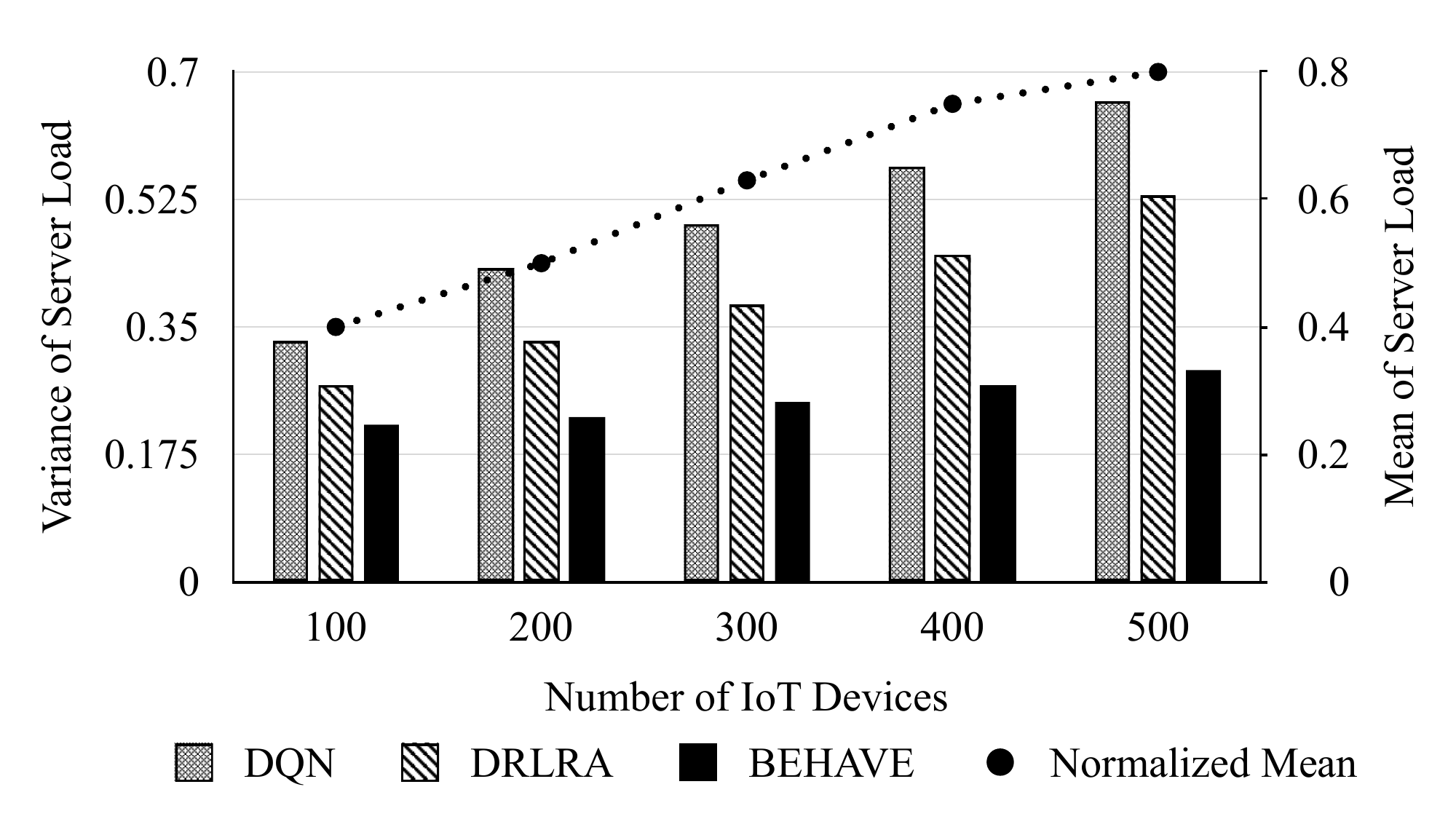}
\vspace{-6 pt}
\caption {\small Variance and mean of edge server load} 
\vspace{-10pt}
\label{load1}
\end{figure}

\begin{figure}[!ht]
\includegraphics[width=.45\textwidth, height=3.5cm]{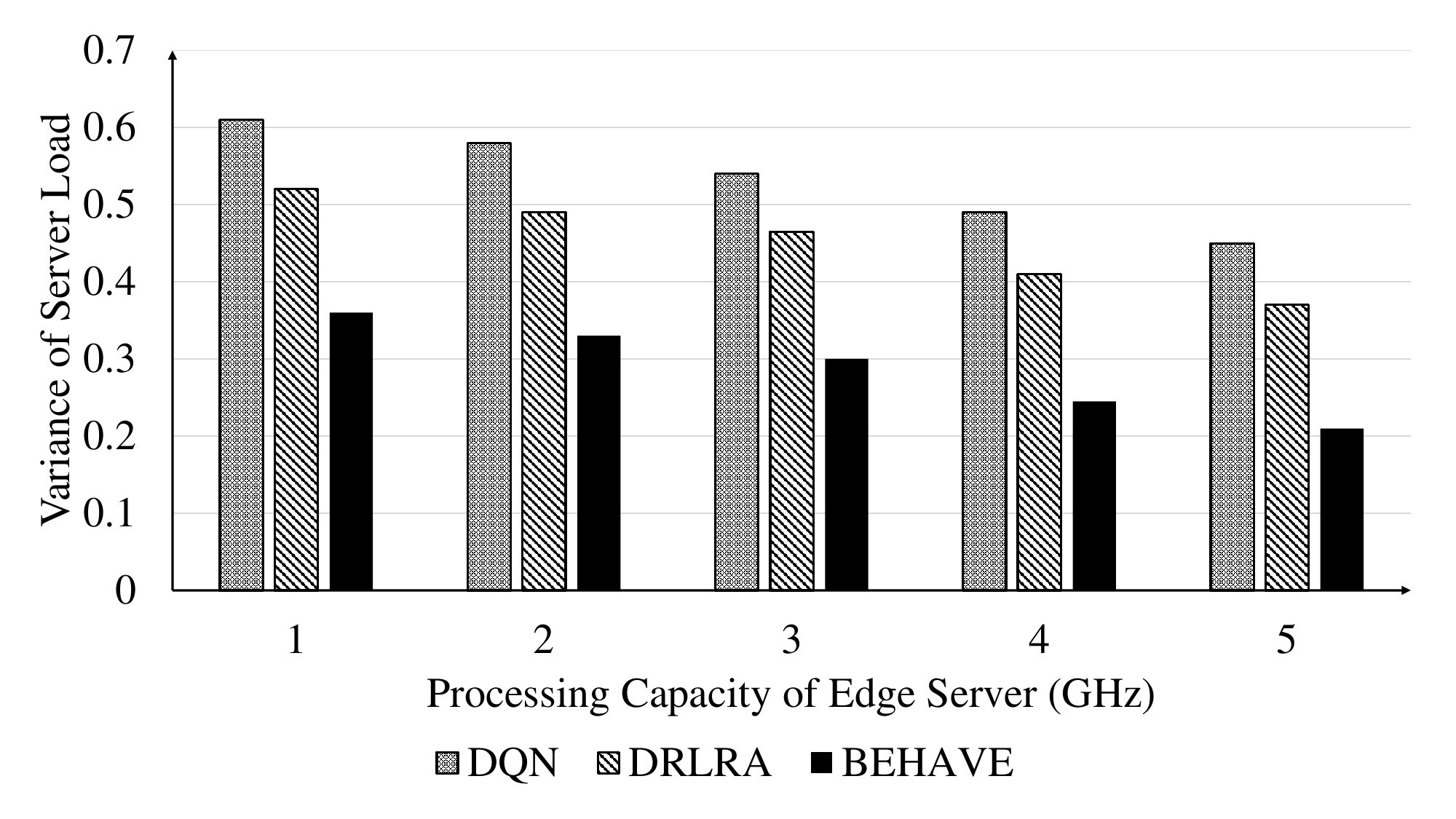}
\vspace{-4mm}
\caption {\small Variance of edge server load vs edge servers' capacity} 
\vspace{-8mm}
\label{load2}
\end{figure}

In this evaluation, we focus on the impact of the number of IoT devices and processing capacity of edge servers on the balance of resource allocation over the edge servers. 
Fig. \ref{load1} shows that as the number of IoT devices grows, the variance of edge servers load increases  
since the number of resource requests increases at each edge server. Moreover, Fig. \ref{load1} presents the normalized mean of the servers load vs. the number of IoT devices. The normalized mean of the servers' load is defined as the ratio between the load of IoT tasks and the capacity of the edge servers and takes a value between 0 and 1. It is noticed that while all the schemes have the same mean of servers load, the variance is different. The mean of servers' load is identical for all schemes as amount of tasks requested to be processed by certain IoT devices is the same and the variance value varies as the amount of the tasks assigned to each edge server (i.e. how much is the load for each edge server) is different.
\textit{BEHAVE} is very effective in balancing the load on edge servers as shown in Fig. \ref{load1} as it makes the variance more stable than other schemes. It is also evident that the merit of \textit{BEHAVE} is clear over other schemes when the number of devices increases.
Fig. \ref{load2} indicates that with increasing of processing capacity of edge servers, the server load variance and the computation load become more manageable and balanced. 
We also notice in Fig. \ref{load2} that
as the processing power increases, \textit{BEHAVE} outperforms other schemes.
\vspace{-4mm}


\subsection{\textit{BEHAVE} Convergence }
\begin{figure}[!ht]
\includegraphics[width=.45\textwidth, height=3.8cm]{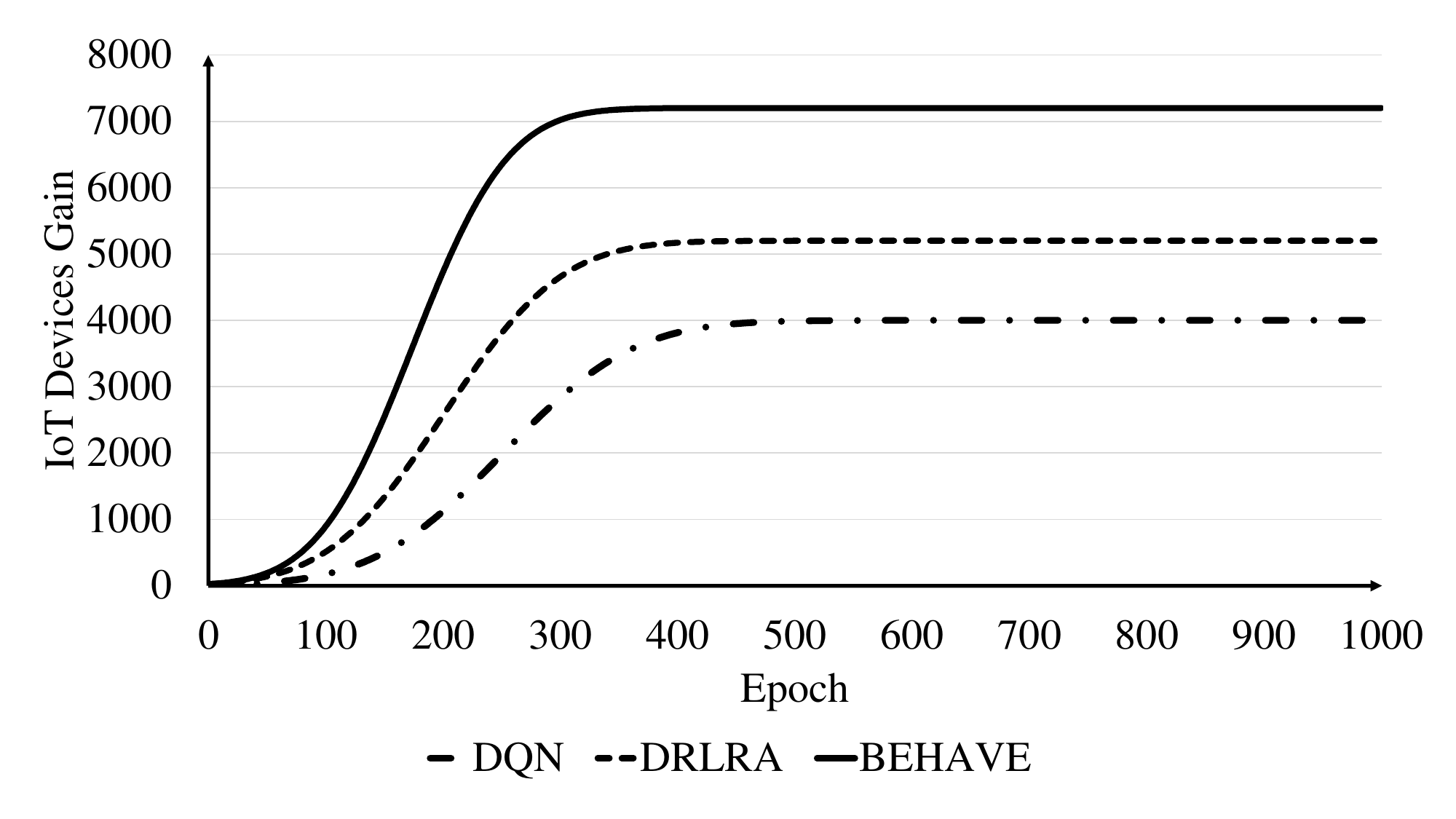}
\vspace{-4mm}
\caption {\small Schemes Convergence} 
\vspace{-3mm}
\label{converge}
\end{figure}

We conduct this evaluation to demonstrate the convergence performance of \textit{BEHAVE}. Fig. \ref{converge} shows the achieved gain of the IoT devices against the number of epoch. We notice that at the beginning, the gain is low because DRL agent does not have enough experience 
 to make rational decisions for resource allocation. With the increase in
the number of epoch, the gain increases
gradually until a relatively stable value is reached. 
\vspace{-4mm}
Fig. \ref{converge} also shows that \textit{BEHAVE} converges faster than other DRL based schemes. 

All the evaluations reveal the advantages of \textit{BEHAVE} design principles including: 1) IoT devices' BRD modeling and assessment in two granularities using novel deep learning scheme; 2) resource allocation model that is characterized as rational, fair and truthful; 3) the enhanced MDP model that collects full state information including: device budget, resource requests and resource availability at all edge servers; and 4) EDRL scheme with novel accurate value function approximation for solving dimensionality problem of DRL.

\vspace{-3mm}
\section{Conclusion}
The paper has tackled the resource allocation problem in Edge-IoT environment with consideration of the IoT devices' BRD. We proposed \textit{BEHAVE} framework which comprises BRD modeling and assessment mechanisms, novel RFTA model and EDRL scheme. The modeling of BRD mechanism exploits deep learning to assess the IoT devices' BRD. 
RFTA model assigns budgets to IoT devices according to their practiced behavior. These budgets are used to obtain resources from the edge that satisfies the IoT devices' demands. RFTA model achieves rational and fair resource allocation policy that encourages IoT devices to be truthful in reporting their resource demands. \textit{BEHAVE} employs an EDRL scheme which learns by reinforcement resource allocation policy that maximizes the devices' gain. Moreover, \textit{BEHAVE} exploits a sophisticated deep neural network to approximate value function in DRL. 
Evaluation results demonstrate \textit{BEHAVE}'s capabilities including detection of devices' irregular BRD, optimizing devices' gain, and maintaining fairness.
\vspace{-0.3cm}
\section*{Acknowledgment}
The work is supported by National Science Foundation (NSF) CNS core grant No. 1909520 and by National Aeronautics and Space Administration (NASA) EPSCoR research grant under No. NNX15AK38A. 
\vspace{-0.3cm}


\begin{IEEEbiography}[{\includegraphics[width=1in,height=1.25in,clip,keepaspectratio]{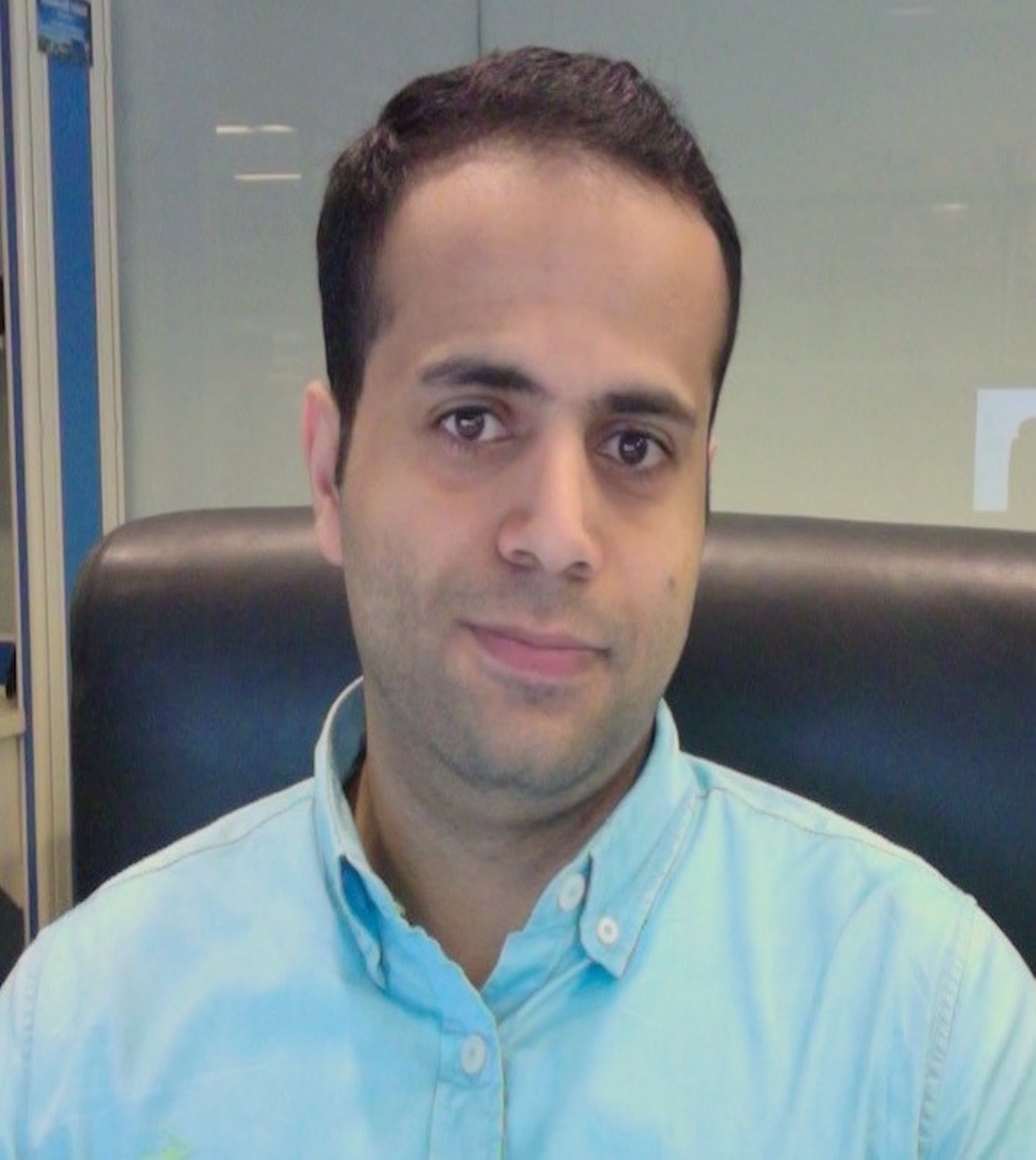}}]{Ismail AlQerm}
is a postdoctoral research associate in the department of computer science at University of Missouri-Saint Louis (UMSL). He received his PhD in computer science from King Abdullah University of Science and Technology (KAUST) in 2017 and was among the recipients of KAUST Provost Award. His research interests include edge computing, resource allocation in IoT networks, developing machine learning techniques for resource allocation in wireless networks, and software defined radio prototypes. He is a member of IEEE and ACM. 
\end{IEEEbiography}

\begin{IEEEbiography}[{\includegraphics[width=1in,height=1.25in,clip,keepaspectratio]{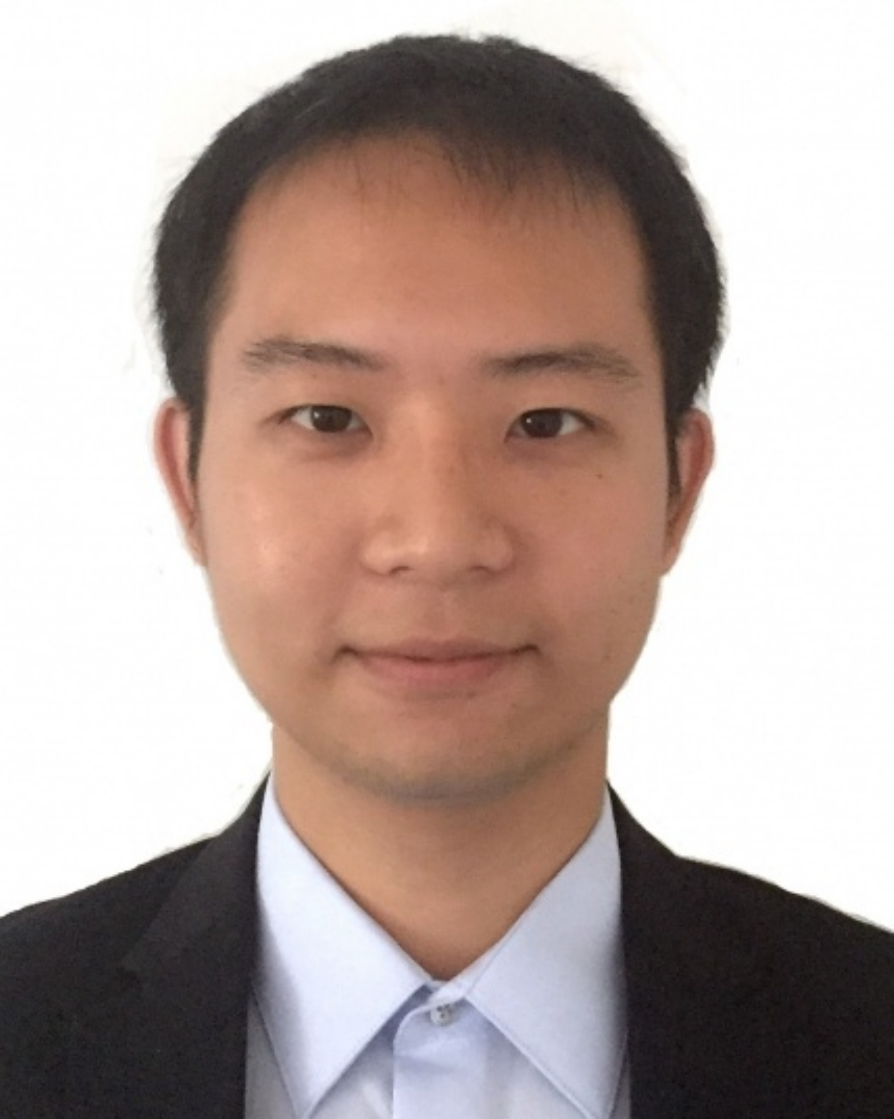}}]{Jianyu Wang}
is currently a Ph.D. student with the Department of Computer Science at the University of Missouri, St. Louis. He received an M.S. in Electrical and Computer Engineering from the Rutgers University, New Brunswick. His current research interests include edge cloud and mobile cloud computing. 
\end{IEEEbiography}


\begin{IEEEbiography}[{\includegraphics[width=1in,height=1.25in,clip,keepaspectratio]{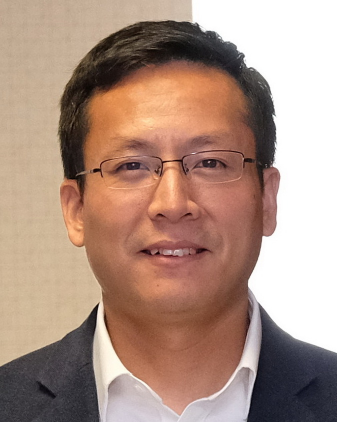}}]{Jianli Pan}
is currently an Associate Professor in the Department of Computer Science at the University of Missouri,
St. Louis, MO USA. He obtained his Ph.D. and M.S. degrees from the Department of
Computer Science and Engineering of Washington University in
St. Louis, USA. He also holds a M.S. degree in
Information Engineering from Beijing University of Posts and
Telecommunications (BUPT), China. He is an associate
editor for both IEEE Communication Magazine and IEEE Access. His
current research interests include Internet of Things (IoT), edge computing, machine learning, cybersecurity, and smart energy. 
\end{IEEEbiography}

\begin{IEEEbiography}[{\includegraphics[width=1in,height=1.25in,clip,keepaspectratio]{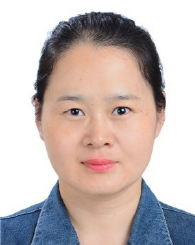}}]{Yuanni Liu}
is an associate professor at the Institute of Future Network Technologies, Chong Qing University of Posts and Telecommunications. She received her Ph.D. from the Department of network technology, Beijing University of Posts and Telecommunications, China, in 2011. Her research interests include mobile crowd sensing, IoT security, and data virtualization.
\end{IEEEbiography}


\begin{thebibliography}{1}


%
\bibitem{PAN18a}
Jianli Pan and James McElhannon, ``Future Edge Cloud and Edge Computing for Internet of Things Applications,'' IEEE Internet of Things Journal, Special Issue on Fog Computing in IoT, Volume: 5, Issue: 1, pp:439-449, ISSN: 2327-4662, DOI: 10.1109/JIOT.2017.2767608, February 2018.







\bibitem{iot2025}
Statista, ``Internet of Things (IoT) connected devices installed base worldwide from 2015 to 2025 (in billions),'' The Statistics Portal, 2019. [Online]. Available:
https://www.statista.com/statistics/471264/iot-number-of-connected-devices-worldwide.


\bibitem{DDOS162}
N.~Woolf, ``DDoS attack that disrupted internet was largest of its kind in history, experts say,'' The Guardian, available at: https://www.theguardian.com/technology/2016/oct/26/ddos-attack-dyn-mirai-botnet.



\bibitem{gridhack18}
Soltan, Saleh, Prateek Mittal, and H. Vincent Poor. ``BlackIoT: IoT botnet of high wattage devices can disrupt the power grid,'' 27th {USENIX} Security Symposium ({USENIX} Security 18). 2018.




\bibitem{CHE16}
Xu Chen, et al., ``Efficient multi-user computation offloading for mobile-edge cloud computing,'' IEEE/ACM Transactions on Networking 5 (2016): 2795-2808.



\bibitem{RAN15}
Rajiv Ranjan, Boualem Benatallah, Schahram Dustdar, and Michael P Papazoglou, ``Cloud resource orchestration programming: Overview, issues, and directions,'' IEEE Internet Computing, 19(5):46-56, 2015.







\bibitem{TRA19}
T. X. Tran and D. Pompili, ``Joint Task Offloading and Resource Allocation for Multi-Server Mobile-Edge Computing Networks,'' in IEEE Transactions on Vehicular Technology, vol. 68, no. 1, pp. 856-868, Jan. 2019.


\bibitem{ZHAN19}
J. Zhang et al., ``Joint Resource Allocation for Latency-Sensitive Services Over Mobile Edge Computing Networks With Caching,'' in IEEE Internet of Things Journal, vol. 6, no. 3, pp. 4283-4294, June 2019.

\bibitem{ZHAO19}
L. Zhao, J. Wang, J. Liu and N. Kato, ``Optimal Edge Resource Allocation in IoT-Based Smart Cities,'' in IEEE Network, vol. 33, no. 2, pp. 30-35, March/April 2019.

\bibitem{KHAL19}
A. Khalili, S. Zarandi and M. Rasti, ``Joint Resource Allocation and Offloading Decision in Mobile Edge Computing,'' in IEEE Communications Letters, vol. 23, no. 4, pp. 684-687, April 2019.

\bibitem{LEG18}
K. Zhang, S. Leng, Y. He, S. Maharjan and Y. Zhang, ``Mobile Edge Computing and Networking for Green and Low-Latency Internet of Things,'' in IEEE Communications Magazine, vol. 56, no. 5, pp. 39-45, May 2018.

\bibitem{ZHA19}
J. Zhang et al., ``Joint Resource Allocation for Latency-Sensitive Services Over Mobile Edge Computing Networks With Caching,'' in IEEE Internet of Things Journal, vol. 6, no. 3, pp. 4283-4294, June 2019.

\bibitem{NIU19}
X. Niu et al., ``Workload Allocation Mechanism for Minimum Service Delay in Edge Computing-Based Power Internet of Things,'' in IEEE Access, vol. 7, pp. 83771-83784, 2019.


\bibitem{LI18}
S. Li et al., ``Joint Admission Control and Resource Allocation in Edge Computing for Internet of Things,'' in IEEE Network, vol. 32, no. 1, pp. 72-79, Jan. 2018.

\bibitem{SUTT98}
R. S. Sutton and A. G. Barto, ``Reinforcement Learning: An Introduction,'' Second edition, The MIT Press, Cambridge, Massachusetts, 1998.






\bibitem{NING19}
Zhaolong Ning, Peiran Dong, Xiaojie Wang, Joel J. P. C. Rodrigues, and Feng Xia, ``Deep Reinforcement Learning for Vehicular Edge Computing: An Intelligent Offloading System,'' ACM Trans. Intell. Syst. Technol. 10, 6, Article 60 (October 2019), 24. 2019. 

\bibitem{YANG18}
T. Yang, Y. Hu, M. C. Gursoy, A. Schmeink and R. Mathar, ``Deep Reinforcement Learning based Resource Allocation in Low Latency Edge Computing Networks,'' 2018 15th International Symposium on Wireless Communication Systems (ISWCS), Lisbon, 2018, pp. 1-5.


\bibitem{ZENG19}
D. Zeng, L. Gu, S. Pan, J. Cai and S. Guo, ``Resource Management at the Network Edge: A Deep Reinforcement Learning Approach,'' in IEEE Network, vol. 33, no. 3, pp. 26-33, May/June 2019.

\bibitem{HUA19}
Liang Huang, Xu Feng, Cheng Zhang, Liping Qian, Yuan Wu,
``Deep reinforcement learning-based joint task offloading and bandwidth allocation for multi-user mobile edge computing,'' in Digital Communications and Networks, Volume 5, Issue 1, 2019, Pages 10-17.

\bibitem{KATO19}
J. Wang, L. Zhao, J. Liu and N. Kato, ``Smart Resource Allocation for Mobile Edge Computing: A Deep Reinforcement Learning Approach,'' in IEEE Transactions on Emerging Topics in Computing, Early access 2019.









\bibitem{THA19}
V. Thangavelu, D. M. Divakaran, R. Sairam, S. S. Bhunia and M. Gurusamy, ``DEFT: A Distributed IoT Fingerprinting Technique,'' in IEEE Internet of Things Journal, vol. 6, no. 1, pp. 940-952, Feb. 2019.










\bibitem{JES18}
Pacheco, Jesus, and Salim Hariri, ``Anomaly behavior analysis for IoT sensors,'' Transactions on Emerging Telecommunications Technologies 29, no. 4 (2018): e3188.



\bibitem{SON19}
Song, Yubo, Qiang Huang, Junjie Yang, Ming Fan, Aiqun Hu, and Yu Jiang, ``IoT device fingerprinting for relieving pressure in the access control,'' In Proceedings of the ACM Turing Celebration Conference-China, p. 143. ACM, 2019.

\bibitem{LAVE18}
Vincent Francois-Lavet, Peter Henderson, Riashat Islam, Marc G. Bellemare, Joelle Pineau, ``An Introduction to Deep Reinforcement Learning,'' in An Introduction to Deep Reinforcement Learning, 2018.

\bibitem{gan}
Goodfellow, Ian, et al., ``Generative adversarial nets,'' Advances in neural information processing systems (NIPS). 2014.

\bibitem{Yaa10}
Asrul H. Yaacob, Ian KT Tan, Su Fong Chien, and Hon Khi Tan, ``Arima based network anomaly detection,'' In 2010 Second International Conference on Communication Software and Networks, pp. 205-209. IEEE, 2010.

\bibitem{Xu19}
Kuai Xu, Yinxin Wan, Guoliang Xue, and Feng Wang, ``Multidimensional behavioral profiling of internet-of-things in edge networks,'' In Proceedings of the International Symposium on Quality of Service (IWQoS 2019). Association for Computing Machinery, New York, NY, USA, Article 37.

\bibitem{AE}
Chong Zhou, and Randy C. Paffenroth, ``Anomaly detection with robust deep autoencoders,'' In Proceedings of the 23rd ACM SIGKDD International Conference on Knowledge Discovery and Data Mining, pp. 665-674. 2017.

\bibitem{ocnn}
Chalapathy, Raghavendra, Aditya Krishna Menon, and Sanjay Chawla, ``Anomaly detection using one-class neural networks,'' arXiv preprint arXiv:1802.06360 (2018).

\bibitem{bigan}
Jeff Donahue, Philipp Krähenbühl, and Trevor Darrell, ``Adversarial feature learning.'' arXiv preprint arXiv:1605.09782 (2016).

\bibitem{lstm}
Sepp Hochreiter and J. Schmidhuber, ``Long Short-Term Memory,'' Neural Comput. 9, 8 (November 1997), 1735–1780.














\bibitem{oc-svm}
Scholkopf, Bernhard, Robert C. Williamson, Alex J. Smola, John Shawe-Taylor, and John C. Platt, ``Support vector method for novelty detection,'' In Advances in neural information processing systems, pp. 582-588. 2000.

\bibitem{BES12}
Ugo Besson, ``The history of the cooling law: when the search for simplicity can be an obstacle,'' Science \& Education 21, no. 8 (2012): 1085-1110.

 \bibitem{MOU04}
H. Moulin, ``Fair Division and Collective Welfare'', Cambridge, MA, USA,
MIT Press, 2004.

\bibitem{PUTE94}
M. L. Puterman, ``Markov Decision Processes: Discrete Stochastic Dynamic Programming,'' New York: Wiley, 1994.

 \bibitem{amazon}
Amazon EC2 Instance Types. [Online]. Available: \url{https://aws.amazon.com/ec2/instance-types}.

 \bibitem{OpenCV}
Open Source Computer Vision Library. [Online]. Available: \url{https://github.com/opencv/opencv-python}

 \bibitem{language}
Steven Bird, Edward Loper and Ewan Klein (2009), ``Natural Language Processing with Python''. O'Reilly Media Inc.

 \bibitem{sense}
Raspberry Pi Sense HAT Library. [Online]. Available: \url{https://github.com/astro-pi/python-sense-hat}.

\bibitem{F1}
Yutaka Sasaki, ``The truth of the F-measure.'' Teach Tutor mater 1, no. 5 (2007): 1-5.




\bibitem{RC}
Peter J. Rousseeuw, and Katrien Van Driessen, ``A fast algorithm for the minimum covariance determinant estimator,'' Technometrics 41, no. 3 (1999): 212-223.

\bibitem{IF}
Fei Tony Liu, Kai Ming Ting, and Zhi-Hua Zhoum ``Isolation forest,'' In 2008 Eighth IEEE International Conference on Data Mining, pp. 413-422. IEEE, 2008.


\bibitem{Sar16}
Sarah M. Erfani, Sutharshan Rajasegarar, Shanika Karunasekera, and Christopher Leckie, ``High-dimensional and large-scale anomaly detection using a linear one-class SVM with deep learning,'' Pattern Recognition 58 (2016): 121-134.





























































\end{thebibliography}
\end{document}